\documentclass{article}

\usepackage{aditi}

\usepackage{style}

\renewcommand{\ForumContactRow}{%
  \begingroup\small\raggedright
    \ifx\ForumEmail\empty\else
      {\color{ForumAccent}\faEnvelope[regular]~}\ %
      \href{mailto:\ForumEmail}{\textcolor{ForumContactText}{\texttt{\ForumEmail}}}\par
      \vspace{\ForumContactGap}%
    \fi
  \endgroup
}

\usepackage{microtype}
\usepackage{xspace}
\usepackage{url}
\usepackage{booktabs}
\usepackage{tabularx}
\usepackage{algorithm}
\usepackage{algorithmic}

\definecolor{skyblue}{RGB}{204,229,255}
\definecolor{darkblue}{rgb}{0, 0, 0.5}
\definecolor{darkgreen}{rgb}{0.0, 0.5, 0.0}
\definecolor{darkred}{rgb}{0.6, 0.0, 0.0}
\definecolor{AuraRow}{RGB}{226,238,255}
\definecolor{KeyCell}{RGB}{255,231,204}

\definecolor{FailText}{RGB}{200,20,20}
\definecolor{BaseRow}{gray}{0.92}

\hypersetup{colorlinks=true, citecolor=darkblue, linkcolor=darkblue, urlcolor=darkblue}

\usepackage[most]{tcolorbox}
\tcbuselibrary{skins,breakable}

\colorlet{TakeawayColor}{green}

\newtcolorbox{takeaway}[1]{
  enhanced, breakable,
  colframe=black!12,
  boxrule=0.35pt,
  arc=1mm,
  title={\textbf{#1}},
  coltitle=black,
  fonttitle=\sffamily\bfseries,
  colbacktitle=TakeawayColor!15!white,
  colback=TakeawayColor!5!white,
  boxed title style={
    sharp corners, boxrule=0pt,
    top=3pt, bottom=3pt, left=4mm, right=4mm,
    borderline={0.5pt}{0pt}{black!10}
  },
  attach boxed title to top left={xshift=4mm, yshift*=-1.2mm},
  boxsep=1.5mm, top=1.5mm, bottom=1.5mm, left=4mm, right=4mm,
  before skip=10pt, after skip=10pt
}

\newtcolorbox{takeawaygreen}[1]{
  enhanced, breakable,
  colframe=black!12, boxrule=0.35pt, arc=1mm,
  title={\textbf{#1}}, coltitle=black, fonttitle=\sffamily\bfseries,
  colbacktitle=green!15!white, colback=green!5!white,
  boxed title style={sharp corners, boxrule=0pt, top=3pt, bottom=3pt, left=4mm, right=4mm, borderline={0.5pt}{0pt}{black!10}},
  attach boxed title to top left={xshift=4mm, yshift*=-1.2mm},
  boxsep=1.5mm, top=1.5mm, bottom=1.5mm, left=4mm, right=4mm,
  before skip=10pt, after skip=10pt
}

\newtcolorbox{takeawayblue}[1]{
  enhanced, breakable,
  colframe=black!12, boxrule=0.35pt, arc=1mm,
  title={\textbf{#1}}, coltitle=black, fonttitle=\sffamily\bfseries,
  colbacktitle=blue!15!white, colback=blue!5!white,
  boxed title style={sharp corners, boxrule=0pt, top=3pt, bottom=3pt, left=4mm, right=4mm, borderline={0.5pt}{0pt}{black!10}},
  attach boxed title to top left={xshift=4mm, yshift*=-1.2mm},
  boxsep=1.5mm, top=1.5mm, bottom=1.5mm, left=4mm, right=4mm,
  before skip=10pt, after skip=10pt
}

\newtcolorbox{takeawayred}[1]{
  enhanced, breakable,
  colframe=black!12, boxrule=0.35pt, arc=1mm,
  title={\textbf{#1}}, coltitle=black, fonttitle=\sffamily\bfseries,
  colbacktitle=red!15!white, colback=red!5!white,
  boxed title style={sharp corners, boxrule=0pt, top=3pt, bottom=3pt, left=4mm, right=4mm, borderline={0.5pt}{0pt}{black!10}},
  attach boxed title to top left={xshift=4mm, yshift*=-1.2mm},
  boxsep=1.5mm, top=1.5mm, bottom=1.5mm, left=4mm, right=4mm,
  before skip=10pt, after skip=10pt
}

\newtcolorbox{takeawayorange}[1]{
  enhanced, breakable,
  colframe=black!12, boxrule=0.35pt, arc=1mm,
  title={\textbf{#1}}, coltitle=black, fonttitle=\sffamily\bfseries,
  colbacktitle=orange!20!white, colback=orange!8!white,
  boxed title style={sharp corners, boxrule=0pt, top=3pt, bottom=3pt, left=4mm, right=4mm, borderline={0.5pt}{0pt}{black!10}},
  attach boxed title to top left={xshift=4mm, yshift*=-1.2mm},
  boxsep=1.5mm, top=1.5mm, bottom=1.5mm, left=4mm, right=4mm,
  before skip=10pt, after skip=10pt
}

\newtcolorbox{takeawaypurple}[1]{
  enhanced, breakable,
  colframe=black!12, boxrule=0.35pt, arc=1mm,
  title={\textbf{#1}}, coltitle=black, fonttitle=\sffamily\bfseries,
  colbacktitle=violet!15!white, colback=violet!5!white,
  boxed title style={sharp corners, boxrule=0pt, top=3pt, bottom=3pt, left=4mm, right=4mm, borderline={0.5pt}{0pt}{black!10}},
  attach boxed title to top left={xshift=4mm, yshift*=-1.2mm},
  boxsep=1.5mm, top=1.5mm, bottom=1.5mm, left=4mm, right=4mm,
  before skip=10pt, after skip=10pt
}

\newtcolorbox{takeawayteal}[1]{
  enhanced, breakable,
  colframe=black!12, boxrule=0.35pt, arc=1mm,
  title={\textbf{#1}}, coltitle=black, fonttitle=\sffamily\bfseries,
  colbacktitle=teal!15!white, colback=teal!5!white,
  boxed title style={sharp corners, boxrule=0pt, top=3pt, bottom=3pt, left=4mm, right=4mm, borderline={0.5pt}{0pt}{black!10}},
  attach boxed title to top left={xshift=4mm, yshift*=-1.2mm},
  boxsep=1.5mm, top=1.5mm, bottom=1.5mm, left=4mm, right=4mm,
  before skip=10pt, after skip=10pt
}

\newtcolorbox{responsebox}[2][]{
    breakable, enhanced,
    colback=white, colframe=blue!50!black, coltext=black, coltitle=white,
    fonttitle=\bfseries\rmfamily, arc=3mm, boxrule=1pt,
    title=#2, #1
}

\newtcolorbox{equationbox}[1]{
  colback=white, colframe=gray!75!black, boxrule=1pt,
  title=#1,
  attach boxed title to top left={yshift=-2mm, xshift=2mm},
  colbacktitle=gray!75!black, coltitle=white, fonttitle=\bfseries\sffamily,
  boxed title style={boxrule=0pt, frame code={}}
}

\usepackage{graphicx}
\usepackage{enumitem}
\usepackage{float}
\newcommand{\reservespace}[1]{\par\ifdim\dimexpr\pagegoal-\pagetotal\relax<#1\relax\newpage\fi}
\usepackage{listings}

\setTitleruleGap{0.75pt}

\title{AURA: Auditory Spatial Intelligence via a Unified Representation
  from a General-Audio Encoder through Complete Alignment}

\setauthors{%
  Boyan Han$^{1}$\thanks{Equal contribution.\quad $\dagger$ Corresponding author.}
  \authorsep
  Xiaoyan Wang$^{1}$\footnotemark[1]
  \authorsep
  Zhe Sun$^{1}$
  \authorsep
  Ziqi Wang$^{2}$
  \authorsep
  Yujun Cai$^{1}$
  \authorsep
  Hongli Cao$^{1}$ \\

  Yiwei Wang$^{1}$
  \authorsep
  Ming-Hsuan Yang$^{3}$
  \authorsep
  Liang An$^{1,\dagger}$
}

\setaffils{%
  $^{1}$Southeast University, \quad
  $^{2}$Rutgers University, \quad
  $^{3}$University of California, Merced
}

\setemail{boyanhan02@gmail.com}

\begin{document}

\maketitle

\begin{abstract}
Answering spatial questions about multi-source scenes requires a model both to identify each sound and to locate it. Existing spatial audio-language models obtain spatial information from a dedicated encoder, which separates spatial and semantic processing into distinct feature spaces. We study whether a general-audio codec trained without spatial supervision can serve both purposes. A controlled layer-wise analysis shows that the frozen codec retains spatial information, with lateral cues carried by the difference between the two ear channels. This information, however, is not readily accessible: in dual-source scenes, a model trained on these features predicts left/right at chance and, on the remaining axes, assigns the detected direction to the queried and the non-queried source equally often. We attribute this failure to an audio--text information asymmetry, in which the audio contains both sources while the answer describes only the queried one. A controlled $2\times2$ experiment shows that either completing the scene description or exposing the inter-channel difference lifts left/right accuracy above chance, whereas binding directions to the queried source requires both. Building on these findings, AURA keeps the codec and the language model frozen and exceeds the strongest dedicated system on SpatialSoundQA by 10.31 points in dual-source direction accuracy, with about 40\% of its trainable parameters.
\end{abstract}

\FloatBarrier
\section{Introduction}
\label{sec:introduction}
Spatial hearing allows humans not only to recognize \emph{what} a sound is but also to infer \emph{where} it originates, estimate its distance, and perceive the spatial relations among multiple sources. Since most real scenes contain several sources at once, these abilities are needed jointly: a listener has to determine which sound comes from where. This capability underlies embodied intelligence, augmented and virtual reality, and assistive hearing. Large audio-language models (LALMs) have advanced rapidly on speech and general audio tasks~\citep{pengi,ltu,salmonn,qwenaudio,qwen2audio,audioflamingo,gama,qwen25omni}, but these advances mainly concern \emph{what} a sound represents rather than \emph{where} it originates.

\begin{figure}[!ht]
    \centering
    \includegraphics[width=\linewidth]{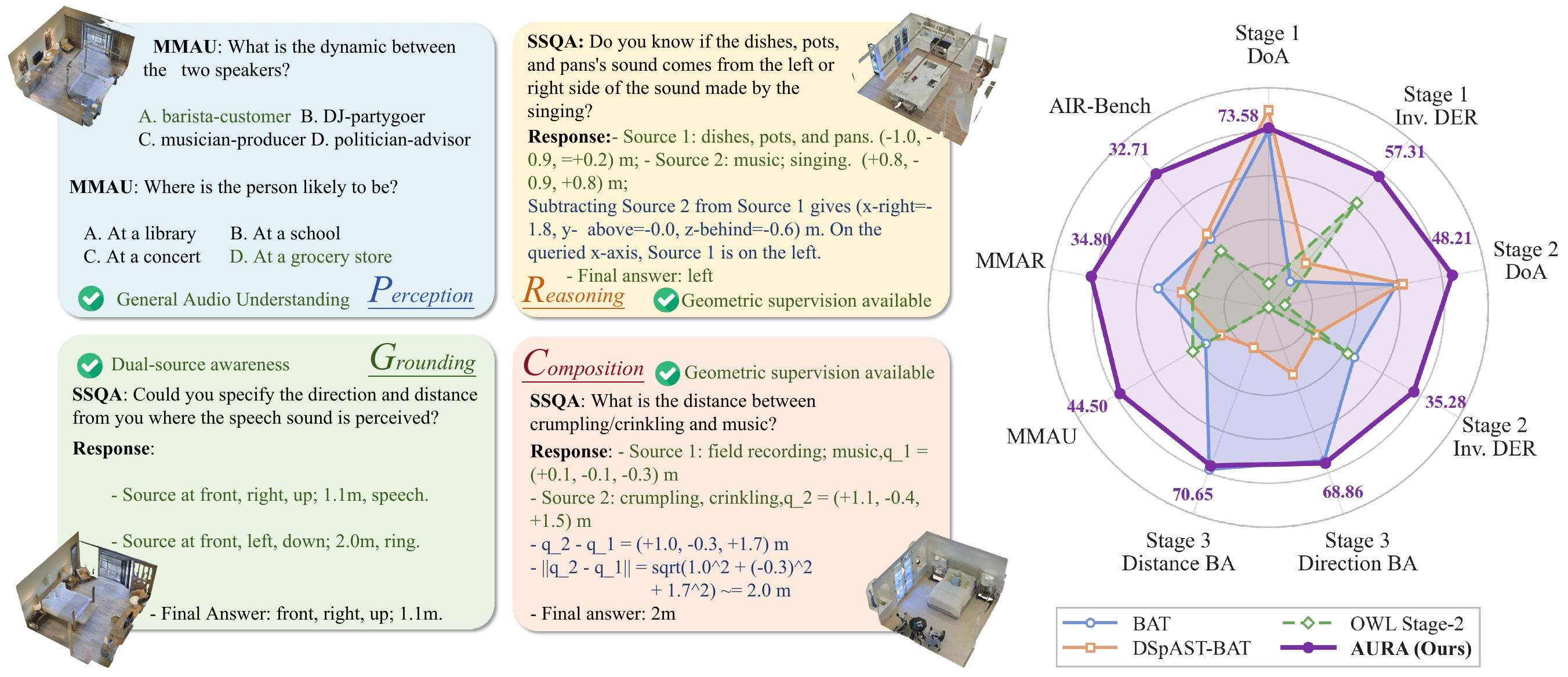}
    \caption{
        \textbf{Spatial and general-audio understanding with AURA.}
        Left: illustrative questions and responses covering general-audio
        understanding, source descriptions, and geometric relations. Source
        coordinates in responses are model outputs. Right: performance overview
        for AURA and the evaluated spatial baselines; AURA attains results
        comparable to dedicated state-of-the-art systems on spatial audio
        understanding and generalizes well to general-audio understanding.
        Matterport3D imagery is subject to the \href{http://kaldir.vc.cit.tum.de/matterport/MP_TOS.pdf}{Matterport3D Terms of Use}.
    }
    \label{fig:capabilities}
\end{figure}

Recent auditory spatial intelligence systems address the second question by adding a dedicated encoder trained for sound localization or geometric perception~\citep{bat,dspastbat,owl}. Spatial and semantic information are then processed in separate representation spaces, and the spatial encoder requires its own pre-training on spatially annotated data. In our evaluation, these systems also fall well behind general-purpose audio-language models on general-audio benchmarks (33.20\% for BAT versus 55.80\% for Qwen2-Audio on MMAU), whereas the general-purpose models, which receive monaural input, perform at or below chance on spatial questions (8.85\% single-source direction accuracy for Qwen2-Audio, against 12.57\% for a random predictor). Spatial and general audio capabilities thus remain largely separate in current systems.

An alternative is to derive spatial information from a general-purpose audio representation. Neural audio codecs are a promising choice for this purpose. Because a codec is trained to reconstruct the waveform, its representation of each channel has to preserve the level and timing detail from which interaural cues arise~\citep{banc,vcnac}. At the same time, a codec trained with the reconstruction objective alone, although it preserves such acoustic detail, yields tokens whose semantic content is weak, which degrades the content accuracy of language models built on them~\citep{xcodec}. X-Codec addresses this by injecting semantic features from a pretrained audio encoder before quantization; it thus retains the acoustic detail of a reconstruction codec while strengthening the semantic content of its tokens, which makes it the best-suited choice among released codecs for a representation that has to both identify and locate sounds. Prior work has shown that codecs preserve binaural structure at the level of the overall representation~\citep{phaseconfounds}, but it remains unclear how this information is distributed across layers and whether a language model can exploit it. We therefore analyze X-Codec~\citep{xcodec}, a general-audio codec trained without spatial supervision, by varying binaural room impulse responses with the source content fixed, and vice versa. Shallow residual vector quantization (RVQ) layers are more sensitive to source content, whereas deeper layers respond increasingly to interaural differences, room acoustics, and reverberation. Lateral information, moreover, is only partially recoverable from either channel alone and becomes reliably decodable through the difference between the two, so it has to be recovered through cross-channel comparison.

The presence of this information, however, does not imply that a model can use it. When the RVQ features of each channel are projected into a large language model (LLM) and trained with question--answer pairs on dual-source SpatialSoundQA~\citep{bat}, left/right prediction remains at chance, while front/back and up/down exceed it. Even the above-chance axes do not reflect source-specific localization. When both sources lie on the same side of the queried axis, the answer does not depend on which source is queried, and the model performs well; when they lie on opposite sides, its accuracy falls to chance. The model thus reads out the spatial attributes of the scene but fails to bind them to the queried source.

We trace this failure to the composition of the training pairs. Each dual-source clip contains two sources, whereas the answer describes only the queried one, leaving the other source without a textual counterpart. We refer to this mismatch as audio--text information asymmetry. Left/right prediction faces an additional obstacle at the interface, which leaves cross-channel comparison to the LLM. Accordingly, we introduce two interventions. On the text side, scene supervision requires the model to describe both sources before answering; on the representation side, an explicit inter-channel difference exposes the lateral cues. In a controlled $2\times2$ experiment, either intervention lifts left/right accuracy well above chance, whereas binding on opposite-side questions improves clearly only when the two are combined.

Building on these findings, we propose \textbf{AURA} (\textbf{A}uditory spatial intelligence via a \textbf{U}nified \textbf{R}epresentation from a general-audio encoder through complete \textbf{A}lignment), which makes the spatial information in a frozen general-audio codec usable through two components, one for each cause identified above. The representation component acts on the audio side. Guided by the layer-wise analysis, the shallow and deep RVQ layers of each channel are aggregated into a content-oriented and an acoustic-oriented band, and the band-wise difference between the left and right channels is added as two further slots. The resulting six-slot representation keeps the original channel features, which carry the front/back, up/down, and distance cues, while presenting the lateral contrast within individual tokens; because the added slots are deterministic functions of the codec features, no spatial encoder or external acoustic feature is introduced. The alignment component acts on the text side. Scene supervision requires the model to describe every audible source, with its category, direction, and distance, before giving the answer, so that each source in the audio has a textual counterpart. Relational questions further require comparing the positions of two sources, which a list of sources does not state; for these questions, geometric supervision inserts an explicit comparison of the two positions between the scene description and the answer. The codec and the Qwen3-8B backbone remain frozen throughout, and only a projector and LoRA modules, 64.63M parameters in total, are trained.

On SpatialSoundQA, AURA performs competitively with dedicated spatial systems (Figure~\ref{fig:capabilities}). Its advantage is largest in dual-source scenes, where it exceeds the strongest dedicated system by 10.31 points in direction accuracy while using about 40\% of that system's trainable parameters; it also scores higher than these systems on MMAU, MMAR, and AIR-Bench. Since AURA and the baselines differ in several components, we interpret these results as evidence that a frozen general-audio codec is sufficient for competitive spatial question answering on SpatialSoundQA, rather than as evidence that it outperforms a dedicated encoder.
\vspace{0.5cm}
\begin{takeaway}{Key Contributions}
\begin{enumerate}
    \item We provide a controlled layer-wise analysis of a general-purpose audio codec trained without spatial supervision, revealing that shallow RVQ layers are more sensitive to source content, while deeper layers respond more strongly to spatial acoustic cues. We further identify inter-channel representational differences associated with left-right spatial information.
    \item We identify previously overlooked failures in spatial information accessibility and source--attribute binding, and propose the information asymmetry hypothesis to explain them. Controlled experiments demonstrate that either scene-aware supervision or explicit binaural differences can restore spatial readout, while substantial binding improvements require their combination.
    \item Building on these findings, we propose AURA, which addresses the two identified causes with two components, a unified representation and complete audio--text alignment, while keeping the codec and the language model frozen. System-level results support the design: with approximately 40\% of the trainable parameters of dedicated systems, AURA exceeds the strongest of them by 10.31 points in dual-source direction accuracy on SpatialSoundQA and scores higher than them on MMAU, MMAR, and AIR-Bench.
\end{enumerate}
\end{takeaway}

\section{Related Work}
\label{sec:related_work}

\paragraph{Spatial audio-language understanding.} Sound event localization and detection has traditionally relied on dedicated multichannel models and datasets \citep{seldnet,starss23}. BAT introduced SpatialSoundQA and pairs a binaural Spatial-AST encoder, built on the Audio Spectrogram Transformer \citep{ast}, with LLaMA-2 \citep{llama2} \citep{bat}; DSpAST learns disentangled spatial representations \citep{dspastbat}, and OWL trains a geometric encoder with depth and room impulse responses \citep{owl}. Sci-Phi trains dual spatial and spectral encoders on first-order Ambisonics and makes enumerating all sources and room parameters its objective \citep{sciphi}; SAIL separates acoustic and spatial streams in a dedicated transformer and aligns them through source-slot queries in a dual-stream Q-Former \citep{sail}. \citet{tang2024spatialaudio} pair a frozen Whisper encoder on the omnidirectional Ambisonics channel with hand-crafted intensity vectors, without which localization fails: spatial processing again resides in a separate feature space, and since the encoder sees one channel, whether a general-purpose encoder retains spatial cues from both channels remains untested. STAR-Bench reports large gaps to human performance on spatial reasoning \citep{starbench}. ELSA and SALM learn spatial audio--text alignment \citep{elsa,salm}; SPUR, Spatial-Omni, and PhaseCoder add spatial pathways to general-purpose models \citep{spur,spatialomni,phasecoder}; multi-object binding and query-conditioned separation have also been studied \citep{worldnotmono,dynamicsource}. These works rely on task-specific inductive biases and separate feature spaces, leaving open whether spatial understanding must rely on a separately trained feature space. AURA shows that a general-audio codec trained without spatial supervision already retains cues for binaural multi-source understanding: its frozen features yield competitive SpatialSoundQA results without a dedicated spatial encoder, and its general-audio understanding generalizes to standard benchmarks.

\paragraph{Codec representations and accessible physical cues.} Neural audio codecs discretize audio with residual vector quantization \citep{soundstream,encodec,dac} and underpin codec language models for generation \citep{audiolm,valle}. SpeechTokenizer guides its first RVQ layer with a HuBERT semantic teacher \citep{hubert} and tests content--speaker separation through voice conversion \citep{speechtokenizer}; TacoLM and ELLA-V build autoregressive and non-autoregressive TTS on the first and residual EnCodec codebooks, improving gated text conditioning and reordering phoneme and audio tokens for alignment, respectively \citep{tacolm,ellav}; ALMTokenizer trains a low-bitrate codec with semantic priors across all three VQ layers \citep{almTokenizer}; X-Codec enriches codec tokens with semantic features for generation \citep{xcodec}. These works study speech content, speaker identity, token predictability, and generation, leaving open whether a general-audio codec trained without spatial supervision retains binaural and room information across its RVQ layers. SonicBench finds physical cues more accessible to frozen-encoder probes than to the complete audio-language model \citep{sonicbench} but does not test how codec-layer spatial cues can serve multi-source QA; SARL probes spatial audio encoders for source- and room-level factors and finds source factors easier to decode \citep{sarl}, but its encoders are trained with spatial input, whereas we analyze a codec never exposed to spatial supervision and ask how its retained cues can be made usable. A related gap between encoded and expressed information has been reported for paralinguistic cues \citep{heardnotheeded}, and psychoacoustic probing suggests binaural models can rely on spectro-temporal interference rather than fine interaural phase \citep{phaseconfounds}. We address both questions: controlled source-content, binaural, and room perturbations show that the codec retains spatial information and reveal its layer-wise distribution, and AURA reorganizes the frozen content- and acoustic-oriented bands with explicit interaural contrast to use that information for multi-source spatial QA.

\paragraph{Multi-source audio--text correspondence.} ELSA and SALM study spatial audio--text alignment \citep{elsa,salm}; OWL uses spatially grounded reasoning targets \citep{owl}; SPUR and The World is Not Mono address multi-speaker attribution and multi-object binding \citep{spur,worldnotmono}; Sci-Phi makes exhaustive scene description the task itself \citep{sciphi}, and SAIL enforces source-level correspondence architecturally through source-slot queries \citep{sail}. Our scene supervision differs in role: it is a component of AURA whose effect we isolate as an intervention, holding the audio representation interface fixed so that it tests the text side of the information-asymmetry hypothesis, the side complementary to the representation-side intervention, while the final answer remains the objective. In direction-conditioned speech extraction, \citet{tang2024spatialaudio} observe that the model may transcribe the non-target source or a mixture, degrading as overlap increases; their target likewise contains only the target speaker's transcript, and the failure is left unexplained. None of these works isolates the supervision mismatch we observe when dual-source audio is paired with a target describing only the queried source, under which our frozen-codec, answer-only baseline falls to chance on left/right localization although the representation retains spatial information. AURA addresses the mismatch from both sides, with scene supervision describing both sources and their geometric relations and explicit interaural contrast helping to distinguish the queried source. Controlled experiments with the same codec, backbone, and training samples distinguish scene-level reading from source-specific binding: either intervention improves left/right reading, and their combination yields substantial binding gains, connecting audio--text correspondence and feature organization to the accessibility of retained spatial information.

\paragraph{Binaural spatial question-answering data.} Trainable binaural spatial question-answering corpora remain scarce. SpatialSoundQA renders AudioSet clips with binaural room impulse responses simulated by SoundSpaces~2.0 in Matterport3D buildings \citep{bat}; BiDepth, introduced with OWL, uses the same sources, simulator, and buildings \citep{owl} and thus offers no independent test of generalization. Most other recent resources use first-order Ambisonics rather than binaural input \citep{starss23,spatialomni,sciphi}, and real binaural recordings such as MRSAudio provide trajectories and transcripts without a question-answering protocol \citep{mrsspeech}. We therefore train and evaluate on SpatialSoundQA, as do BAT and DSpAST-BAT.

\FloatBarrier
\section{Preliminaries}
\label{sec:preliminaries}

\subsection{Binaural Spatial Cues}
\label{sec:preliminaries:cues}

Humans rely on two primary categories of acoustic cues to perceive sound source direction \citep{blauert}. Binaural cues are interaural time differences (ITDs), arising from differences in arrival time between the two ears, and interaural level differences (ILDs), caused mainly by the acoustic shadow of the head. They vary predominantly with the lateral position of a source and are the primary basis for left-right localization; because they are defined by differences between the two ear signals, they cannot be extracted from either channel alone and require cross-channel comparison. Spectral cues arise from the direction-dependent filtering of the pinnae, head, and torso and are particularly important for resolving directions with similar binaural cues, such as front-back and up-down ambiguities; unlike binaural cues, they are present in each ear's signal and can be extracted from individual channels. Source distance is primarily inferred from the received sound level and the direct-to-reverberant energy ratio.

\subsection{Neural Audio Codecs and Residual Vector Quantization}
\label{sec:preliminaries:rvq}

Neural audio codecs compress waveforms into discrete token sequences from which a decoder reconstructs the waveform \citep{soundstream,encodec,dac}. Because these tokens can be modeled by language models like text tokens, codecs have become a widely adopted representation for audio generation and audio language modeling \citep{audiolm,valle}.

A typical codec consists of an encoder, a residual vector quantizer (RVQ), and a decoder. The encoder maps a waveform $\mathbf{x}$ to latent vectors $\mathbf{V}=E(\mathbf{x})=[\mathbf{v}_1,\ldots,\mathbf{v}_T]$. At each time step $t$, RVQ approximates $\mathbf{v}_t$ progressively with $N$ codebooks: at the $i$-th layer, the codeword $\mathbf{c}_{t,i}$ closest to the current residual is selected from codebook $\mathcal{C}_i$ and subtracted, and the remainder is passed to the next layer:
\begin{equation}
\label{eq:rvq}
\mathbf{c}_{t,i}=\underset{\mathbf{w}\in\mathcal{C}_i}{\arg\min}\,\left\|\mathbf{r}_{t,i}-\mathbf{w}\right\|_2^2,\qquad
\mathbf{r}_{t,i+1}=\mathbf{r}_{t,i}-\mathbf{c}_{t,i},\qquad
\mathbf{r}_{t,0}=\mathbf{v}_t.
\end{equation}

The quantized vector is the sum of the selected codewords, $\hat{\mathbf{v}}_t=\sum_{i=0}^{N-1}\mathbf{c}_{t,i}$. Since each layer encodes only the residual left by preceding layers, different layers may capture distinct aspects of the signal; in speech codecs, earlier layers have been observed to capture mainly semantic content and deeper layers finer acoustic details such as speaker characteristics \citep{speechtokenizer}.

\subsection{Binaural Scene Construction and Evaluation}
\label{sec:preliminaries:ssqa}

SpatialSoundQA (SSQA) \citep{bat} constructs binaural scenes by taking AudioSet clips \citep{audioset} as sources and simulating binaural room impulse responses (BRIRs) with SoundSpaces 2.0 \citep{soundspaces2} in Matterport3D environments \citep{matterport3d}. For a scene with $M\in\{1,2\}$ static sources, where source $a_m$ at position $\mathbf{p}_m$ has BRIR $h_m=(h_m^L,h_m^R)$, the two channels are rendered by convolution:
\begin{equation}
\label{eq:render}
x^{q}=\sum_{m=1}^{M}a_m*h_m^{q},\qquad q\in\{L,R\},
\end{equation}
where $*$ denotes convolution. In two-source scenes, the sources belong to distinct sound categories and are superimposed within each channel.

SSQA evaluates direction and distance perception in single- and two-source scenes, as well as binary questions about spatial relations between sources; in two-source scenes, the queried source is specified by its sound category. Directions are defined in a listener-centered coordinate system along three axes: left/right, front/back, and up/down. Following SSQA \citep{bat}, we report Direction-of-Arrival Accuracy (DoA), which requires correct predictions on all three axes simultaneously; Distance Error Rate (DER), the proportion of predictions whose absolute distance error exceeds $0.5$~m; and Binary Accuracy (BA) on binary spatial-relation questions, reported separately for direction and distance relations.

\FloatBarrier
\section{Spatial Information in a General-Audio Codec}
\label{sec:rvq_analysis}

Before asking whether a language model can use spatial information from a general-audio codec, we need to know whether that information is present in the codec's representations and where it lies. We analyze the general-audio model of X-Codec \citep{xcodec}. As a reconstruction codec it preserves the acoustic detail of each channel, and the semantic features it distills from a HuBERT teacher adapted to general audio supply the content information that purely acoustic codecs lack (Section~\ref{sec:introduction}), so a single representation can serve both halves of the task. Because different RVQ layers can have distinct representational preferences (Section~\ref{sec:preliminaries:rvq}), we measure the sensitivity of its RVQ layers to spatial changes layer by layer, holding one factor fixed and varying another (Figure~\ref{fig:rvq_layer_analysis}): the two ears of the same rendering (interaural differences), the same source under different impulse responses (room), the same impulse response with different sources (content), and full versus truncated impulse responses (reverberation). Section~\ref{sec:rvq_analysis:similarity} defines the measure and reports these controls, Section~\ref{sec:rvq_analysis:probe} tests whether the spatial attributes they point to can be read out linearly, and together they inform the interface examined in Section~\ref{sec:information_asymmetry}.

\subsection{Controlled Analysis of Layer-wise RVQ Representations}
\label{sec:rvq_analysis:similarity}
\begin{figure}[p]
    \centering
    \includegraphics[width=0.99\linewidth]{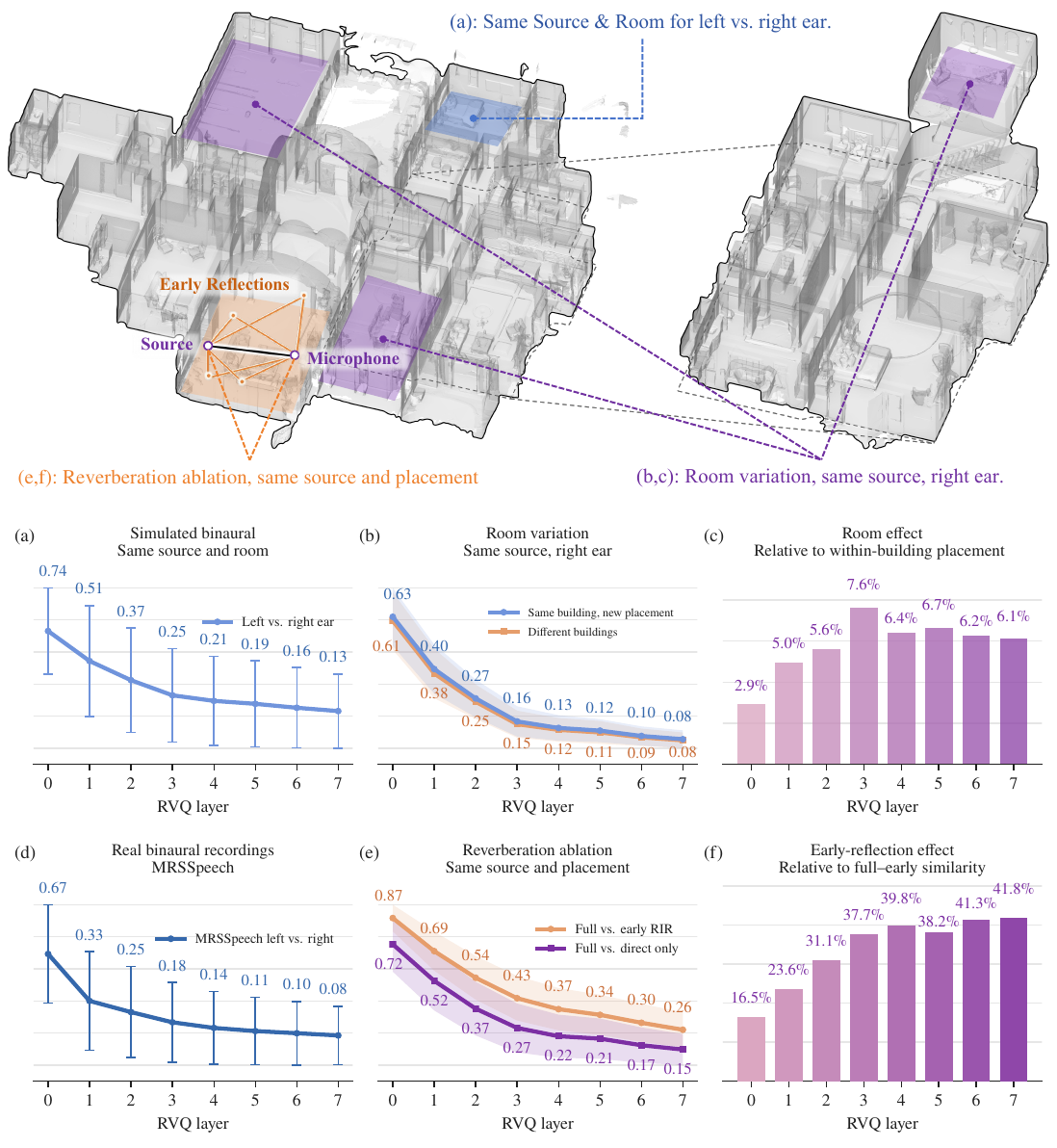}
    \caption{
        \textbf{Layer-wise similarity analysis of RVQ representations.}
        Top: schematic illustration of acoustic variables on Matterport3D scene \texttt{wc2JMjhGNzB}.
        (a,d)~Interaural similarity for simulated (fixed source and BRIR) and real (MRSSpeech) binaural audio;
        the RVQ vectors diverge markedly as the layer index increases.
        (b,c)~Room variation: right-ear similarity when the same audio is rendered at another placement
        in the same building or with an impulse response from a different building; the similarity of the
        RVQ vectors still decreases clearly with depth.
        (e,f)~Full versus early-reflection-only and direct-path-only impulse responses for the same
        placement, and the additional reduction from removing early reflections.
        These comparisons reveal increasing sensitivity to spatially relevant acoustic changes in deeper RVQ layers.
        Scene data under the \href{http://kaldir.vc.cit.tum.de/matterport/MP_TOS.pdf}{Matterport3D Terms of Use}.
    }
    \label{fig:rvq_layer_analysis}
\end{figure}
The analysis uses the general-audio X-Codec encoder with eight RVQ layers at $50$~Hz, each with a codebook of $1024$ $1024$-dimensional embeddings; the encoder, RVQ modules, and codebooks remain frozen throughout the analysis and all downstream training. Inputs are single-source binaural audio from SpatialSoundQA, rendered from a source $a$ and a BRIR $h=(h^{L},h^{R})$ as in Eq.~\eqref{eq:render} with $M=1$. For two rendered inputs with codeword embeddings $\mathbf{c}_{t,i}^{(1)}$ and $\mathbf{c}_{t,i}^{(2)}$ at aligned time step $t$ and RVQ layer $i$, the layer-wise similarity is
\begin{equation}
\operatorname{sim}_{i}
=\frac{1}{T}\sum_{t=1}^{T}
\frac{\mathbf{c}_{t,i}^{(1)}\cdot\mathbf{c}_{t,i}^{(2)}}
{\|\mathbf{c}_{t,i}^{(1)}\|_2\|\mathbf{c}_{t,i}^{(2)}\|_2},
\qquad i\in\{0,\ldots,7\}.
\end{equation}

We first hold the source and BRIR fixed and compare the left and right channels of the same rendering, which measures sensitivity to interaural differences since the source is identical. Next, we hold the source fixed and change the BRIR, comparing the same channel across placements within one building and across buildings; conversely, we fix the BRIR and vary the source to measure the effect of content changes under the same acoustic response. For a fixed source and placement, we also compare full, early-reflection-only, and direct-path-only responses to examine reverberation sensitivity. Finally, we repeat the left--right analysis on real binaural recordings from the MRSSpeech subset of MRSAudio \citep{mrsspeech} to check whether the layer-wise trend persists outside convolution-based simulation.

Figure~\ref{fig:rvq_layer_analysis} illustrates the acoustic variables and summarizes the comparisons. Interaural similarity in simulated binaural audio decreases from $0.742$ at L0 to $0.130$ at L7 (panel~a), and real recordings show the same trend, from $0.674$ to $0.083$ (panel~d), so the pattern is not an artifact of convolutional rendering.

For fixed source content, changing the impulse response across buildings reduces similarity more than changing placement within a building (panels~b,c), with the relative reduction growing from approximately $2.9\%$ at L0 to $6$--$8\%$ across L3--L7; under a fixed source and placement, removing early reflections yields an additional relative reduction growing from approximately $16\%$ at L0 to $42\%$ at L7 (panels~e,f). Deeper layers are thus more sensitive to environmental acoustic variation.

The complementary fixed-BRIR, varying-source control (not shown in the figure) yields similarities of $0.316$ at L0, $0.111$ at L2, and $0.040$ at L3, approaching the random-code reference in deeper layers. Together, the controls indicate that shallow layers are more strongly influenced by source content, whereas deeper residual layers are increasingly sensitive to interaural, room, and reverberation changes; this empirical division motivates the band aggregation used below without implying that either group exclusively represents one factor. The interaural and placement trends also appear on two reconstruction-only codecs (Appendix~\ref{app:codecs}).

\subsection{Linear Decodability of Spatial Attributes}
\label{sec:rvq_analysis:probe}

The controlled comparisons of Section~\ref{sec:rvq_analysis:similarity} show that the frozen codec's layer-wise representations respond systematically to binaural, room, and reverberation changes, with distinct preferences across RVQ layers. Sensitivity, however, does not imply that spatial attributes are explicitly encoded or directly recoverable, so we use linear probing to test whether they can be decoded from the frozen RVQ representations.

\begin{wraptable}{r}{0.56\linewidth}
    \vspace{-0.6\baselineskip}
    \centering
    \small
    \setlength{\tabcolsep}{3.2pt}
    \caption{
        Linear probes of spatial attributes from the frozen RVQ
        representation. The linear classifiers clearly exceed the chance-level
        reference, showing that the RVQ representation carries information for
        understanding the spatial relations in audio. Binaural audio further
        outperforms monaural audio, most markedly on left/right discrimination,
        where the second channel removes most of the remaining error
        (26.0\% to 7.0\%), showing that the interaural cross-channel
        difference is what makes left/right reliably decodable. Reference rows
        are constant predictions without audio.
    }
    \label{tab:probe_readability}
    \begin{tabular*}{\linewidth}{@{\extracolsep{\fill}}lccccc@{}}
        \toprule
        & \multicolumn{4}{c}{\textbf{Direction} $\uparrow$}
        & \textbf{Dist.} \\
        \cmidrule(lr){2-5} \cmidrule(l){6-6}
        \textbf{Features}
        & DoA & L/R & F/B & U/D
        & MAE $\downarrow$ \\
        \midrule
        \rowcolor{black!8}
        \multicolumn{6}{@{}l}{\textit{Single source}} \\
        Reference
        & 15.6 & 50.9 & 50.9 & 59.0 & 1.249 \\
        Monaural
        & 39.5 & 74.0 & 71.7 & 72.1 & 1.051 \\
        Binaural
        & 51.1 & 93.0 & 72.9 & 73.4 & 1.029 \\
        \midrule
        \rowcolor{black!8}
        \multicolumn{6}{@{}l}{\textit{Dual source (queried source)}} \\
        Reference
        & 15.4 & 50.4 & 49.8 & 55.5 & 1.695 \\
        Monaural
        & 23.8 & 65.4 & 61.6 & 59.8 & 1.574 \\
        Binaural
        & 28.1 & 72.3 & 64.0 & 60.9 & 1.575 \\
        \bottomrule
    \end{tabular*}
\end{wraptable}

The probes operate on layer-wise vectors retrieved from the frozen codebooks. Guided by the preferences observed in Section~\ref{sec:rvq_analysis:similarity}, we sum the codeword vectors of layers L0--L2 into a content-oriented band and of L3--L7 into an acoustic-oriented band, and average each band over the entire recording to obtain a fixed-length representation. To examine how attributes are distributed across channels, we compare monaural features, the two bands of a single channel, with binaural features, the corresponding bands of both channels concatenated. Separate ridge classifiers are trained for the three directional axes and a ridge regressor for distance (in meters); regularization is selected on a held-out 10\% subset of the training audio, after which the probes are refitted on the full training set. Probes are trained and evaluated separately for single-source scenes (278,665 training recordings, evaluated on the complete set of 29,651 recordings) and dual-source scenes (236,000 training questions, evaluated on 10,640 directional questions with the queried source as the target), with no overlap between training and evaluation audio. Reference baselines are constant predictions without audio input: the majority class for direction and the training-set mean for distance.

Table~\ref{tab:probe_readability} presents the results. First, spatial attributes are linearly decodable from the frozen codec representations. In single-source scenes, all attributes substantially outperform their constant baselines: with binaural features, left-right, front-back, and up-down accuracies reach 93.0\%, 72.9\%, and 73.4\%, joint accuracy across the three axes reaches 51.1\% against a baseline of 15.6\%, and distance mean absolute error (MAE) decreases from 1.249~m to 1.029~m. Directional and distance information can thus be recovered by linear readout from a frozen general-purpose codec trained without spatial supervision. Second, the distribution of decodable attributes is consistent with established binaural cue theory. Front-back and up-down accuracies already reach 71.7\% and 72.1\% with monaural features and improve only modestly to 72.9\% and 73.4\% with the second channel, suggesting that spectral cues are largely preserved within individual channels, whereas left-right accuracy rises substantially from 74.0\% to 93.0\%. The benefit of combining both channels is therefore concentrated on the left-right axis: lateral information is recovered more effectively through inter-channel comparison than from either channel alone. Third, spatial information remains decodable in dual-source scenes, but its extraction becomes considerably more challenging. With the queried source as the target, binaural features reach 72.3\%, 64.0\%, and 60.9\% on the left-right, front-back, and up-down axes and a joint accuracy of 28.1\%, above the 15.4\% reference, but left-right accuracy drops from 93.0\% to 72.3\% and joint accuracy from 51.1\% to 28.1\% relative to single-source scenes. Since both sources are superimposed within each channel, source-specific spatial information remains present but becomes substantially harder to extract from the scene-level representations. Appendix~\ref{app:mrsspeech_probe} reports the same probes on real binaural recordings.

Importantly, linear decodability establishes the presence of recoverable information but does not guarantee its accessibility to a language model or its correct binding to the queried source; whether these emerge depends on the representation interface and supervision strategy, which the next section investigates.

\FloatBarrier
\section{The Information-Asymmetry Hypothesis}
\label{sec:information_asymmetry}

\subsection{Selective Collapse in Multi-Source Scenes}
\label{sec:information_asymmetry:collapse}
Section~\ref{sec:rvq_analysis} establishes that directional, distance, and content information is linearly decodable from the frozen codec, with lateral information carried mainly by the contrast between the two channels. This section shows that a conventional interface nevertheless fails on exactly that axis, and that the failure has a second component: directional attributes are not bound to the queried source. The interface sums the codeword vectors $\mathbf{c}_{t,i}^{(q)}$ of each channel $q\in\{L,R\}$ into the two bands of Section~\ref{sec:rvq_analysis},
\begin{equation}
\mathbf{e}_{t}^{(q,S)}=\sum_{i=0}^{2}\mathbf{c}_{t,i}^{(q)},
\qquad
\mathbf{e}_{t}^{(q,A)}=\sum_{i=3}^{7}\mathbf{c}_{t,i}^{(q)},
\label{eq:rvq_bands}
\end{equation}
where $S$ and $A$ label the content- and acoustic-oriented bands; the four slots $[L_S,L_A,R_S,R_A]$ are temporally pooled and mapped independently into the language model by a shared projector $g(\cdot)$, and under answer-only supervision the target contains only the final answer about the queried source.

\begin{figure}[!t]
    \centering
    \includegraphics[width=\linewidth]{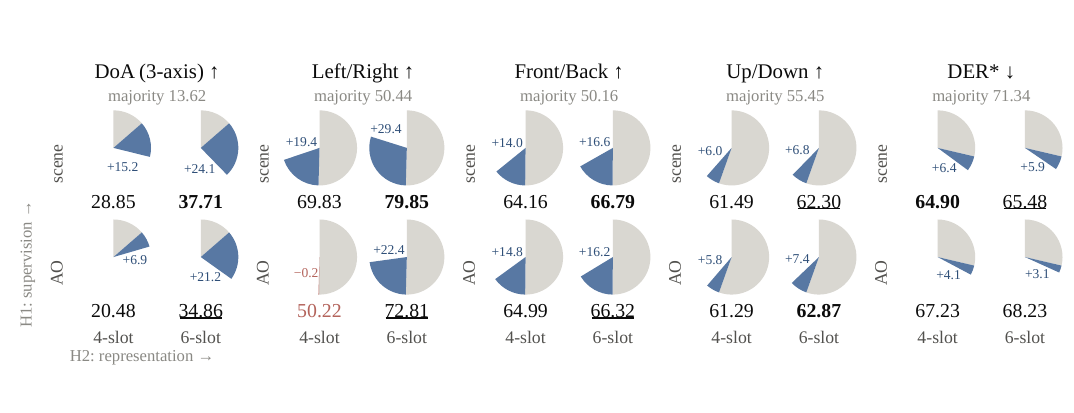}
    \caption{
        Overall results of the $2\times2$ controlled experiment on the
        dual-source direction questions. The four-slot answer-only baseline
        (red) fails to exceed the majority baseline on left/right
        discrimination, whereas every other intervention clearly surpasses the
        baseline level. The best value for each metric is in bold and the
        second best is underlined; the majority baseline is not ranked. H1:
        scene supervision; H2: six-slot representation with interaural
        difference slots; AO: answer-only supervision. Each pie spans 100
        points: the grey sector is the majority-baseline level, the blue sector
        the gain over it, and the number below the pie the metric value (\%);
        for DER ($*$) lower is better, so the pie shows $100-\mathrm{DER}$.
    }
    \label{fig:verification_2x2}
\end{figure}
Trained on dual-source SpatialSoundQA, this interface collapses on exactly one axis (Figure~\ref{fig:verification_2x2}): on the $10{,}640$ dual-source direction questions, front/back and up/down accuracies reach $65.0\%$ and $61.3\%$, above the majority-class baselines of $50.2\%$ and $55.5\%$, whereas left/right accuracy is $50.2\%$, matching the $50.4\%$ majority baseline, with $76.9\%$ of predictions being ``right.'' The only axis at chance is the one carried by interaural differences.

The above-chance axes do not imply source-specific understanding. Stratifying questions by whether the two sources lie on the same or opposite sides of the queried axis (Figure~\ref{fig:reading_binding}) separates two abilities: on the same side the answer is independent of which source is queried, so same-side accuracy measures \emph{reading} the scene-level attribute, whereas on the opposite side the predicted attribute must be associated with the queried source, so opposite-side accuracy measures \emph{binding}. The four-slot baseline reads but does not bind: front/back and up/down reach $80.4\%$ and $77.1\%$ on same-side questions but $50.5\%$ and $49.2\%$ on opposite-side questions, and its full three-axis prediction matches the queried source and the other source equally often (about $20.5\%$ each). Left/right fails even on same-side questions ($49.5\%$), because lateral information requires comparing the two channels (Section~\ref{sec:rvq_analysis:probe}), which this interface leaves to the language model. The failure thus has two components: inaccessible left/right information and unbound directional attributes.

\subsection{The Information-Asymmetry Hypothesis and Two-Sided Interventions}
\label{sec:information_asymmetry:interventions}
The two components of this failure have different origins. The left/right collapse is a property of the interface: Section~\ref{sec:rvq_analysis} showed that lateral information is recovered reliably only through the contrast between the two channels, and the four-slot interface presents the channels as separate tokens and leaves the comparison to the language model. The failure to bind attributes to the queried source points instead to the training signal. For sources $a_1,a_2$ at positions $\mathbf{p}_1,\mathbf{p}_2$, the audio-side scene is
\begin{equation}
\mathcal{I}_{\mathrm{audio}}
=\{(a_1,\mathbf{p}_1),(a_2,\mathbf{p}_2)\},
\end{equation}
whereas answer-only supervision for a question about source $k$ covers only that source ($\bar{k}$ denotes the other):
\begin{equation}
\mathcal{I}_{\mathrm{text}}^{\mathrm{ao}}
=\{(a_k,\mathbf{p}_k)\}\subsetneq\mathcal{I}_{\mathrm{audio}}.
\end{equation}
Coverage here means source identity and the queried spatial attributes, not the waveform or the complete position. Both sources contribute to the audio features, but only one has a textual counterpart, and we hypothesize that this \emph{information asymmetry} impedes alignment between the queried source and its description: the unpaired source contributes interfering features, weakening source--attribute binding and, with it, the supervision from which any cross-channel comparison would have to be learned. The hypothesis motivates an intervention on the text side; the interface finding above motivates a separate one on the representation side. We state both before testing them jointly.

\paragraph{Text-side intervention (H1): completing the scene description.} We supply the missing correspondence by requiring the model to describe both audible sources before answering. With $\mathcal{D}(a_m,\mathbf{p}_m)$ a textual description of a source's category and location and $\mathbf{Y}_{\mathrm{ao}}$ the original answer, the target becomes
\begin{equation}
\mathbf{Y}_{\mathrm{scene}}
=\bigl[\mathcal{D}(a_1,\mathbf{p}_1),\,
\mathcal{D}(a_2,\mathbf{p}_2),\,\mathbf{Y}_{\mathrm{ao}}\bigr],
\qquad
\mathcal{I}_{\mathrm{text}}^{\mathrm{scene}}=\mathcal{I}_{\mathrm{audio}},
\end{equation}
where the equality denotes coverage of both sources. This \emph{scene supervision} gives each source a textual counterpart, so the model must state what each source is and where it is before answering, directly supervising identity--location correspondence.

\paragraph{Representation-side intervention (H2): explicit interaural contrast.} This intervention follows from Section~\ref{sec:rvq_analysis} rather than from the asymmetry: lateral information lies in the contrast between the two channels, which the four-slot interface never forms, and in a dual-source scene each channel token additionally mixes the features of both sources. We expose the contrast directly by retaining the original channels and adding
\begin{equation}
\mathbf{e}^{(S)}=\tfrac12\bigl(\mathbf{e}^{(L)}-\mathbf{e}^{(R)}\bigr).
\label{eq:interaural_contrast}
\end{equation}
Subtracting the channel features suppresses their shared component and places their difference within a single slot, making interaural contrast directly available to the projector instead of leaving the language model to recover it across separate channel tokens. The difference is a deterministic function of the frozen features: it changes their presentation without adding information or isolating an individual source.

\subsection{Controlled Test of the Hypothesis}
\label{sec:information_asymmetry:verification}
\begin{figure}[!t]
    \centering
    \includegraphics[width=\linewidth]{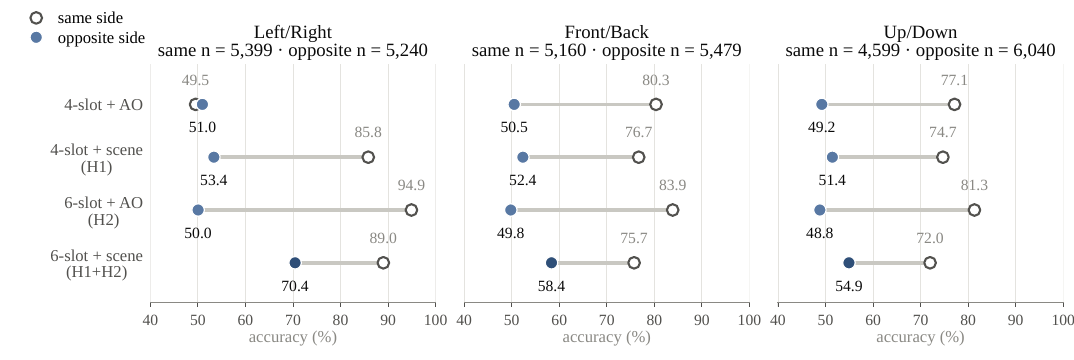}
    \caption{
        Dual-source direction accuracy stratified by whether the non-queried
        source lies on the same side as the queried source along each axis.
        Except for the full intervention, the direction gains of every
        intervention are confined to the same-side stratum (open markers); only
        the full intervention effectively improves the opposite-side questions
        (filled markers), which require binding the semantic and acoustic
        attributes to the queried source, lifting them above the
        majority-baseline level. $n$ is the number of questions per stratum. H1:
        scene supervision; H2: six-slot representation; AO: answer-only
        supervision.
    }
    \label{fig:reading_binding}
\end{figure}
We test the two interventions in a $2\times2$ experiment crossing four- and six-slot representations (the latter adding one difference slot per band) with answer-only and scene supervision. All four runs share the frozen codec, the Qwen3-8B backbone, the projector and LoRA configuration, the training samples, and the optimization procedure; we evaluate with greedy decoding on the dual-source test set and compare matched predictions with paired exact McNemar tests. The text-side factor isolates the variable targeted by the hypothesis: the answer-only target of these direction questions already specifies the full direction and distance of the queried source, so scene supervision adds nothing about it beyond its category, which the question names, and the added content is exactly the description of the non-queried source.

Either intervention alone lifts left/right above chance (Figure~\ref{fig:verification_2x2}). Scene supervision raises the four-slot left/right accuracy from $50.2\%$ to $69.8\%$, the six-slot representation reaches $72.8\%$ under answer-only supervision (both $p<10^{-150}$ against the baseline), and their combination reaches $79.9\%$, significantly above either alone ($p<10^{-48}$); three-axis DoA accuracy rises from $20.5\%$ to $28.9\%$ and $34.9\%$ for the individual interventions and to $37.7\%$ for their combination.

Reading is restored by either intervention; binding requires both. The stratified results (Figure~\ref{fig:reading_binding}) separate the two effects: on same-side questions, left/right accuracy rises from $49.5\%$ to $85.8\%$ with scene supervision and to $94.9\%$ with the six-slot representation, whereas on opposite-side questions scene supervision alone and the six-slot representation alone reach only $53.4\%$ and $50.0\%$ and their combination reaches $70.4\%$, an interaction of $+18.0$ percentage points (bootstrap $95\%$ confidence interval $[15.5,20.4]$). Front/back and up/down improve more moderately, from $50.5\%$ and $49.2\%$ to $58.4\%$ and $54.9\%$ under the combined intervention, with interactions of $+6.7$ ($[4.8,8.7]$) and $+3.9$ ($[2.0,5.8]$) points. Binding thus benefits from both explicit source-wise textual correspondence and explicit interaural contrast, most on the axis carried by interaural differences.

Two further observations sharpen the interpretation. On same-side questions the six-slot answer-only model scores higher than the combined intervention (left/right $94.9\%$ versus $89.0\%$), but a scene-level readout is correct there regardless of which source is queried; scene supervision replaces this shortcut with source-specific answers, which is what produces the binding gains on opposite-side questions. The recovery of reading also does not depend on how the channels are distinguished at the interface: with the identical four-slot interface, a shared projector, and no slot-type embeddings, scene supervision alone raises same-side left/right accuracy to $85.8\%$, so positional information suffices to tell the left and right tokens apart.

The generated descriptions are consistent with the final answers: in both scene-supervised runs, more than $99.9\%$ of outputs contain two parseable sources, which we align to the annotated sources by label overlap; the per-source left/right accuracy of the descriptions rises from $70.3\%$ with four slots to $80.4\%$ with six, and the three-axis accuracy for the queried source ($38.6\%$) is consistent with the final-answer DoA of $37.7\%$.

These results support the information-asymmetry hypothesis and distinguish recovering a directional attribute from associating it with the queried source: either intervention substantially improves left/right reading, but the pronounced binding gains require their combination, which motivates AURA's joint use of scene supervision and the six-slot representation.

\FloatBarrier
\section{AURA}
\label{sec:aura_method}
\begin{figure}[!t]
    \centering
    \includegraphics[width=\linewidth]{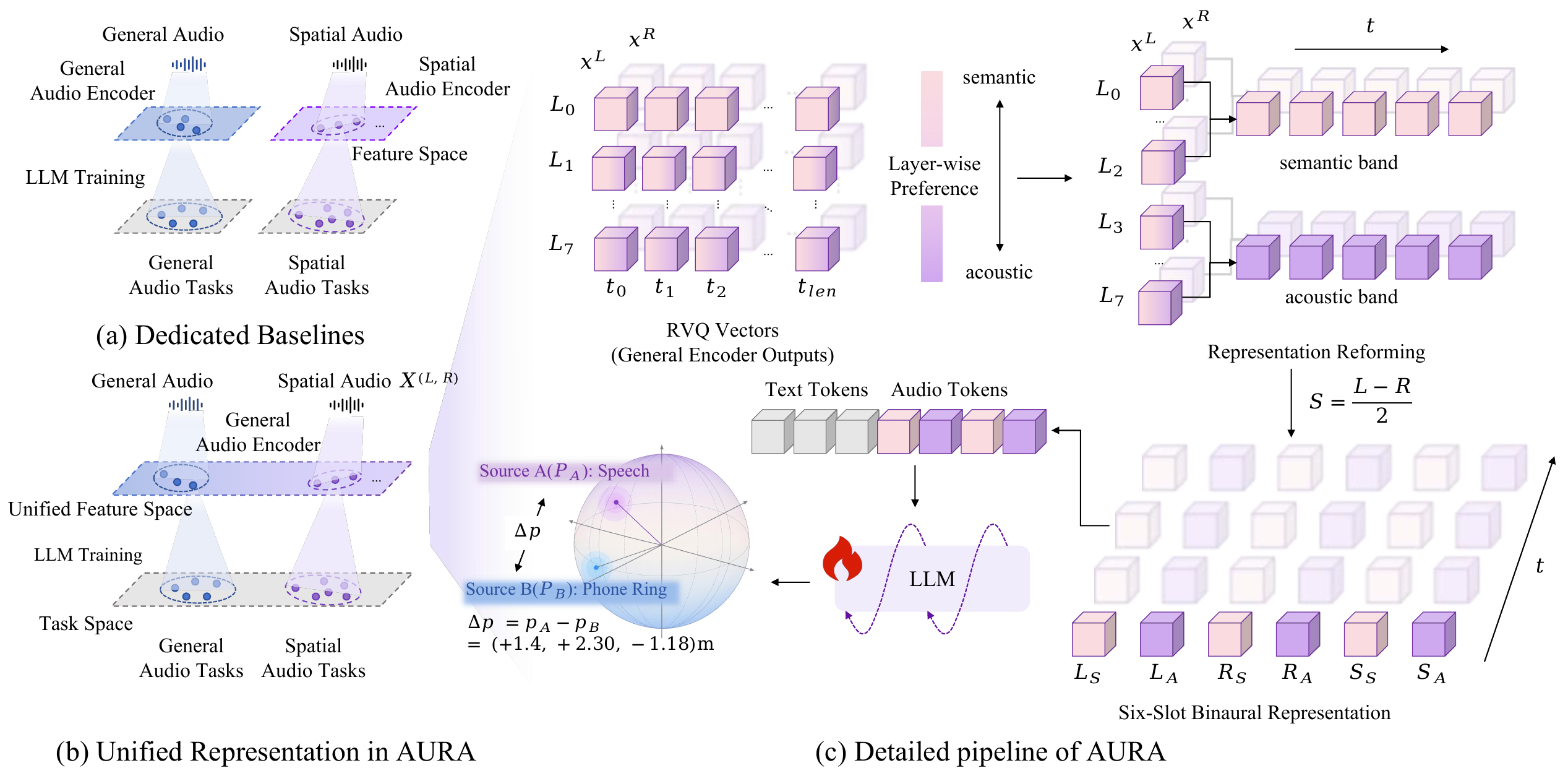}
    \caption{
        \textbf{AURA architecture.} (a) Existing dedicated spatial pathways
        and general-audio pathways are separate and rely on different feature
        spaces, so features cannot be shared for unified perception. (b) AURA
        provides a shared, frozen general-audio codec representation that
        unifies perception in a single audio feature space. (c) Details of the
        AURA pipeline: for each ear, RVQ layers L0--L2 and L3--L7 are summed
        into content-oriented ($S$) and acoustic-oriented ($A$) bands, and
        adding the band-wise interaural difference yields six slots. A shared
        projector maps these slots to audio tokens for the language model,
        which generates a scene description and, for relation questions, a
        geometric analysis before the final answer.
    }

    \label{fig:pipeline}
\end{figure}

Figure~\ref{fig:pipeline} gives an overview of AURA. The following subsections describe its six-slot binaural representation, its scene and geometric supervision, and its two-stage training objective.

\subsection{Six-Slot Binaural Representation}
\label{sec:aura_method:representation}
Given a binaural waveform $\mathbf{x}=(\mathbf{x}^{L},\mathbf{x}^{R})$, each channel is encoded independently by the same frozen encoder $E(\cdot)$ and its $N$-layer RVQ; following Section~\ref{sec:preliminaries:rvq}, the selected codeword vectors at frame $t$ are
\begin{equation}
\bigl[\mathbf{c}_{t,0}^{(q)},\ldots,\mathbf{c}_{t,N-1}^{(q)}\bigr]
=\mathrm{RVQ}\bigl(E(\mathbf{x}^{q})_t\bigr),
\qquad q\in\{L,R\},\quad t=1,\ldots,T,
\end{equation}
where $\mathbf{c}_{t,i}^{(q)}\in\mathbb{R}^{d}$, with $N=8$, $d=1024$, and a frame rate of $50$~Hz. The encoder and codebooks remain frozen, and AURA uses the selected codeword vectors without the codec decoder.

Aggregating the shallow and deep layers into the content- and acoustic-oriented bands of Eq.~\eqref{eq:rvq_bands} gives four slots per frame,
\begin{equation}
\mathbf{F}_t^{(4)}=
\bigl[\mathbf{e}_{t}^{(L,S)},\mathbf{e}_{t}^{(L,A)},
      \mathbf{e}_{t}^{(R,S)},\mathbf{e}_{t}^{(R,A)}\bigr],
\end{equation}
abbreviated as $[L_S,L_A,R_S,R_A]$; these are the four-slot features of Section~\ref{sec:information_asymmetry}.

For each band, we add an explicit interaural difference,
\begin{equation}
\mathbf{e}_{t}^{(S,b)}=
\tfrac12\bigl(\mathbf{e}_{t}^{(L,b)}-\mathbf{e}_{t}^{(R,b)}\bigr),
\qquad b\in\{S,A\},
\label{eq:band_difference}
\end{equation}
so that each frame has six slots,
\begin{equation}
\mathbf{F}_t^{(6)}=
\bigl[\mathbf{e}_{t}^{(L,S)},\mathbf{e}_{t}^{(L,A)},
      \mathbf{e}_{t}^{(R,S)},\mathbf{e}_{t}^{(R,A)},
      \mathbf{e}_{t}^{(S,S)},\mathbf{e}_{t}^{(S,A)}\bigr],
\end{equation}
abbreviated as $[L_S,L_A,R_S,R_A,S_S,S_A]$, where the first index identifies the left, right, or difference stream and the second the content- or acoustic-oriented band ($S_A$ is thus the acoustic-oriented difference slot). The added slots are deterministic functions of the original features and introduce no external acoustic features or spatial supervision, and retaining L/R preserves the original features used to predict front/back, up/down, and distance.

To shorten the input sequence, the $T$ frames are partitioned into consecutive, non-overlapping windows $\mathcal{W}_1,\ldots,\mathcal{W}_J$ of at most $w$ frames, $J=\lceil T/w\rceil$; content-oriented slots use element-wise max pooling to retain prominent responses, and acoustic-oriented slots use mean pooling to aggregate the local acoustic representation:
\begin{equation}
\widetilde{\mathbf{e}}_{j}^{(q,S)}
=\max_{t\in\mathcal{W}_{j}}\mathbf{e}_{t}^{(q,S)},
\qquad
\widetilde{\mathbf{e}}_{j}^{(q,A)}
=\frac{1}{|\mathcal{W}_{j}|}\sum_{t\in\mathcal{W}_{j}}\mathbf{e}_{t}^{(q,A)},
\quad q\in\{L,R,S\}.
\end{equation}
Pooling excludes padded frames, and the difference slots are constructed \emph{before} pooling and pooled with the same band-specific scheme as L/R.

A shared two-layer MLP projector \citep{llava} maps each pooled slot independently into the language model's embedding space,
\begin{equation}
\mathbf{z}_{j}^{(q,b)}=g\bigl(\widetilde{\mathbf{e}}_{j}^{(q,b)}\bigr),
\end{equation}
and the resulting tokens are concatenated in temporal order with a fixed slot order within each window, giving the four- and six-slot sequences
\begin{equation}
\begin{aligned}
\mathcal{Z}_{4}&=
\bigl(\mathbf{z}_{j}^{(L,S)},\mathbf{z}_{j}^{(L,A)},
\mathbf{z}_{j}^{(R,S)},\mathbf{z}_{j}^{(R,A)}\bigr)_{j=1}^{J},\\
\mathcal{Z}_{6}&=
\bigl(\mathbf{z}_{j}^{(L,S)},\mathbf{z}_{j}^{(L,A)},
\mathbf{z}_{j}^{(R,S)},\mathbf{z}_{j}^{(R,A)},
\mathbf{z}_{j}^{(S,S)},\mathbf{z}_{j}^{(S,A)}\bigr)_{j=1}^{J},
\end{aligned}
\end{equation}
of lengths $4J$ and $6J$. AURA prepends $\mathcal{Z}_6$ to the question text as the language-model input.

\subsection{Scene Supervision}
\label{sec:aura_method:supervision}
Consider $M\in\{1,2\}$ sources, where source $a_m$ at position $\mathbf{p}_m$ has BRIR $h_m=(h_m^L,h_m^R)$; up to a shared mixture gain, the rendered audio is
\begin{equation}
\mathbf{x}^{q}=\sum_{m=1}^{M}a_m*h_m^{q},\qquad q\in\{L,R\},
\end{equation}
where $*$ denotes convolution, and its scene inventory is $\mathcal{I}_{\mathrm{audio}}=\{(a_m,\mathbf{p}_m)\}_{m=1}^{M}$. Scene supervision requires the model to describe this inventory before answering: each description $\mathcal{D}(a_m,\mathbf{p}_m)$ contains category labels, three directional attributes relative to the listener, and distance, and the target concatenates the source count, source descriptions, and original answer,
\begin{equation}
\mathbf{Y}_{\mathrm{scene}}=
\bigl[M,\,\mathcal{D}(a_1,\mathbf{p}_1),\ldots,
\mathcal{D}(a_M,\mathbf{p}_M),\,\mathbf{Y}_{\mathrm{ao}}\bigr].
\end{equation}
Sources are ordered by their spatial attributes independently of the question (left before right, then front before behind, then above before below, then increasing distance); this canonical order only resolves the permutation ambiguity of describing an unordered set of sources, is a property of the training target, and does not appear in the input. Single- and dual-source samples share this format, and the controlled experiment in Section~\ref{sec:information_asymmetry} uses the $M=2$ variant without the source-count line.

For questions about relations between two sources, a geometric analysis $\mathcal{G}(\mathbf{p}_1,\mathbf{p}_2)$ is inserted between the scene inventory and the answer:
\begin{equation}
\mathbf{Y}_{\mathrm{geo}}=
\bigl[M,\,\mathcal{D}(a_1,\mathbf{p}_1),\,
\mathcal{D}(a_2,\mathbf{p}_2),\,
\mathcal{G}(\mathbf{p}_1,\mathbf{p}_2),\,\mathbf{Y}_{\mathrm{ao}}\bigr].
\end{equation}
Here each source description additionally contains its three-dimensional listener-centered coordinates, by which sources are ordered lexicographically, again independently of the question; depending on the question, the geometric analysis compares positions along an axis, compares radial distances, or computes the relative displacement between sources. Figure~\ref{fig:target_examples} illustrates both target formats.
\begin{figure}[t]
\centering
\begin{tcolorbox}[enhanced, colback=white, colframe=black!35, boxrule=0.5pt, arc=1mm,
    title={Scene supervision $\mathbf{Y}_{\mathrm{scene}}$}, coltitle=black, colbacktitle=black!8,
    fonttitle=\sffamily\bfseries\small, left=2mm, right=2mm, top=1mm, bottom=1mm]
\ttfamily\footnotesize\raggedright
{\color{black!55}Question: Where is the sound of the battle cry coming from?}\\[2pt]
Scene description:\\
Number of sources: 2\\
- Source at left, front, and below, approximately 5 meters away: music; sampler.\\
- Source at right, behind, and below, approximately 3 meters away: battle cry; cheering; crowd.\\
Final answer: right, behind, below; 3m
\end{tcolorbox}
\vspace{-2pt}
\begin{tcolorbox}[enhanced, colback=white, colframe=black!35, boxrule=0.5pt, arc=1mm,
    title={Geometric supervision $\mathbf{Y}_{\mathrm{geo}}$}, coltitle=black, colbacktitle=black!8,
    fonttitle=\sffamily\bfseries\small, left=2mm, right=2mm, top=1mm, bottom=1mm]
\ttfamily\footnotesize\raggedright
{\color{black!55}Question: Are you able to detect the orchestra's sound coming from the right and the cricket's sounds from the left?}\\[2pt]
Scene description:\\
Number of sources: 2\\
- Source 1: cricket; insect. Listener-centered coordinates (x-right=-1.2, y-above=-1.1, z-behind=+2.6) m; direction: left, below, and behind; radial distance: 3.1 m.\\
- Source 2: orchestra. Listener-centered coordinates (x-right=+4.5, y-above=-1.2, z-behind=+2.1) m; direction: right, below, and behind; radial distance: 5.1 m.\\
Geometric analysis:\\
Source 2 is right on the x-axis, while Source 1 is left on the x-axis. The stated directions are consistent.\\
Final answer: Yes
\end{tcolorbox}
\caption{
    AURA training-target examples for scene supervision (top) and geometric
    supervision (bottom). The question is shown in gray; the remaining text is
    the model's target.
}
\label{fig:target_examples}
\end{figure}

\subsection{Two-Stage Training Objective}
\label{sec:aura_method:objective}
Given audio tokens $\mathcal{Z}_6$, question tokens $\mathbf{P}$, and target $\mathbf{Y}=(y_1,\ldots,y_{|\mathbf{Y}|})$, either $\mathbf{Y}_{\mathrm{scene}}$ or $\mathbf{Y}_{\mathrm{geo}}$, the model learns $p_\theta(\mathbf{Y}\mid\mathcal{Z}_6,\mathbf{P})$ autoregressively, and both training stages minimize the same token-level cross-entropy,
\begin{equation}
\mathcal{L}^{(s)}=-\sum_{n=1}^{|\mathbf{Y}|}
\log p_{\theta_s}\bigl(y_n\mid\mathcal{Z}_6,\mathbf{P},y_{<n}\bigr),
\qquad s\in\{\mathrm{align},\mathrm{joint}\}.
\end{equation}
Alignment optimizes only the projector, $\theta_{\mathrm{align}}=\theta_g$, and joint training optimizes the projector and LoRA \citep{lora} parameters, $\theta_{\mathrm{joint}}=\{\theta_g,\theta_{\mathrm{LoRA}}\}$. The loss is computed only on target tokens, with audio and question tokens masked, and the codec encoder, codebooks, and language-model backbone remain frozen in both stages.

\section{Experiments}
\label{sec:experiments}
\subsection{Experimental Setup}
\label{sec:experiments:setup}
\paragraph{Setting.} SpatialSoundQA (SSQA) \citep{bat} provides binaural scenes with one or two static sources, rendered with BRIRs simulated in Matterport3D rooms. It is the standard trainable binaural spatial question-answering corpus: the alternatives either share its sources and simulator or lack a question-answering protocol (Section~\ref{sec:related_work}), and the dedicated systems we compare with are trained on it. The codec operates at $16$~kHz, so dry clips and BRIRs are both processed at this rate before rendering.

\begin{table*}[t]
    \centering
    \caption{
        \textbf{Main results: system-level comparison on SpatialSoundQA.}
        AURA performs comparably to current dedicated spatial audio
        understanding systems on SpatialSoundQA and achieves the best results
        on single-source DER, dual-source DoA and DER, and the relative-direction
        relation questions. All spatial metrics are reported in percentage
        points. The first group lists references: the random predictor, the
        majority-class baseline, and the zero-shot general-purpose models. The
        best result in each column is in bold and the second best is underlined;
        reference rows are not ranked.
    }
    \label{tab:ssqa_main}

    \small
    \setlength{\tabcolsep}{4.5pt}
    \renewcommand{\arraystretch}{1.08}

    \begin{tabularx}{\linewidth}{lc*{6}{>{\centering\arraybackslash}X}}
        \toprule
        \textbf{Model}
        & \textbf{Params.}
        & \multicolumn{2}{c}{\textbf{Single Source}}
        & \multicolumn{2}{c}{\textbf{Dual Source}}
        & \multicolumn{2}{c}{\textbf{Relation}} \\
        \cmidrule(lr){3-4}
        \cmidrule(lr){5-6}
        \cmidrule(lr){7-8}

        &
        &
        \textbf{DoA} $\uparrow$
        & \textbf{DER} $\downarrow$
        & \textbf{DoA} $\uparrow$
        & \textbf{DER} $\downarrow$
        & \textbf{Dir. BA} $\uparrow$
        & \textbf{Dist. BA} $\uparrow$ \\
        \midrule

        Random
        & --
        & 12.57 & 67.33
        & 12.41 & 67.46
        & 50.00 & 50.00 \\
        Majority baseline
        & --
        & 15.57 & 49.31
        & 13.62 & 71.34
        & 50.00 & 50.67 \\
        Qwen2-Audio
        & --
        & 8.85 & 80.03
        & 9.74 & 80.27
        & 49.32 & 46.14 \\

        Qwen2.5-Omni
        & --
        & 10.61 & 69.50
        & 12.02 & 79.26
        & 49.55 & 54.30 \\
        \midrule
        BAT
        & 161.04M
        & 72.28 & 64.55
        & 36.64 & \underline{70.98}
        & \underline{68.54} & \textbf{71.35} \\

        DSpAST-BAT$^{\dagger}$
        & 161.16M
        & \textbf{80.84} & 60.73
        & \underline{37.90} & 75.06
        & 58.07 & 47.72 \\

        OWL (Stage 1/2)
        & 161.04M
        & 9.82 & \underline{48.17}
        & 13.36 & 71.68
        & \multicolumn{1}{c}{--} & \multicolumn{1}{c}{--} \\

        \midrule

        \rowcolor{AuraRow}
        \textbf{AURA}
        & \textbf{64.63M}
        & \underline{73.58}
        & \textbf{42.69}
        & \textbf{48.21}
        & \textbf{64.72}
        & \textbf{68.86}
        & \underline{70.65} \\

        \bottomrule
    \end{tabularx}
\end{table*}
\paragraph{Implementation details.} Our frozen front end is \texttt{xcodec\_hubert\_general\_audio\_v2} \citep{xcodec}, whose semantic branch builds on HuBERT \citep{hubert}; it operates at $50$~Hz with eight RVQ layers of $1024$ codewords of dimension $1024$. With pooling window $w=4$, a $10$-second input yields $750$ audio tokens. The projector is a two-layer MLP with GELU activation \citep{gelu}. Qwen3-8B \citep{qwen3} is frozen, with rank-$16$ LoRA modules \citep{lora} in its attention and feed-forward layers, giving $64.63$M trainable parameters ($20.98$M in the projector and $43.65$M in LoRA). Training follows a progressive curriculum \citep{curriculum}, each stage initialized from its predecessor and trained for one epoch with AdamW \citep{adamw}, a cosine schedule with warmup, a learning rate of $10^{-4}$, and a global batch size of $16$ ($32$ for projector alignment).

\paragraph{Baselines.} We compare against specialized spatial models, BAT \citep{bat}, DSpAST-BAT \citep{dspastbat}, and OWL \citep{owl}, and general-purpose audio-language models. BAT uses its released checkpoint; as no complete DSpAST-BAT checkpoint is available, we train a local reconstruction on SSQA for one epoch with its released encoder, denoted DSpAST-BAT$^{\dagger}$. OWL is evaluated with its released checkpoints, inference code, and prompt format on the SSQA evaluation samples, using the Stage-1 checkpoint for single-source questions and the Stage-2 checkpoint for dual-source questions; as only its first two training stages are released, its relation-task results are not reported. It answers direction with clock-face hours, which we map to an azimuth of $30^\circ$ per hour, assigning left/right by the sign of its sine and front/back by the sign of its cosine (non-negative values map to right and front), reading the up/down token as above/below, and taking the trailing number as the distance. Its outputs are then scored with the same direction and distance rules as the other models. Qwen2-Audio \citep{qwen2audio} and Qwen2.5-Omni \citep{qwen25omni} represent general-purpose audio-language and omni-modal models; both are evaluated zero-shot with a fixed left-$30^\circ$ monaural beam derived from the binaural input and listed as references, together with the random predictor from BAT and a majority-class baseline that answers the majority class on each direction axis and the best constant distance on the test set.

\begin{wraptable}{r}{0.48\linewidth}
    \vspace{-0.6\baselineskip}
    \centering
    \small
    \setlength{\tabcolsep}{3pt}
    \caption{
        General-audio accuracy (\%). AURA generalizes well to general-audio
        understanding and outperforms the current spatial understanding
        systems. The two general-purpose models are zero-shot references
        trained with large-scale audio instruction data and are not ranked;
        among the spatial systems, the best result is in bold and the second
        best underlined.
    }
    \label{tab:general_audio}
    \begin{tabular}{lccc}
        \toprule
        \textbf{Model}
        & \textbf{MMAU} $\uparrow$
        & \textbf{MMAR} $\uparrow$
        & \textbf{AIR-Bench} $\uparrow$ \\
        \midrule
        Qwen2-Audio
        & 55.80 & 40.00 & 48.95 \\
        Qwen2.5-Omni
        & 65.20 & 49.20 & 53.89 \\
        \midrule
        BAT
        & 33.20 & \underline{31.10} & 26.52 \\
        DSpAST-BAT$^{\dagger}$
        & 31.30 & 29.80 & \underline{26.99} \\
        OWL (Stage 2)
        & \underline{34.90} & 29.20 & 25.38 \\
        \midrule
        \rowcolor{AuraRow}
        \textbf{AURA}
        & \textbf{44.50} & \textbf{34.80} & \textbf{32.71} \\
        \bottomrule
    \end{tabular}
\end{wraptable}
\paragraph{Benchmarks and evaluation metrics.} Spatial understanding is evaluated on SSQA with the metrics of Section~\ref{sec:preliminaries:ssqa}: DoA accuracy (all three axes correct), DER (distance error above $0.5$~m), and BA on the binary relation questions, reported separately for direction and distance relations. SSQA also scores the sound classes named in an answer by a classification mAP that matches the generated text to the $355$ AudioSet class names in a shared text-embedding space \citep{bat}; we do not report it, since it measures closed-set label recall through an embedding-based matcher rather than semantic understanding of the audio. Semantic understanding is instead evaluated by multiple-choice accuracy on MMAU test-mini \citep{mmau}, MMAR base \citep{mmar}, and AIR-Bench Foundation \citep{airbench}: each candidate answer is scored by its mean conditional token log-likelihood, the highest-scoring option is selected, and scores are averaged over non-overlapping $10$-second windows of the full audio.

\subsection{Main Results}
\paragraph{Spatial audio understanding.} Table~\ref{tab:ssqa_main} compares the models on SSQA. With a fully frozen general-audio codec pretrained without spatial supervision, AURA matches or exceeds the specialized spatial systems on four of the six metrics and ranks second on the other two, while the two general-purpose models remain near the random and majority references on direction tasks under the zero-shot, monaural-input setting. The comparison is system-level; the roles of AURA's own components are established by the controlled experiment of Section~\ref{sec:information_asymmetry}.

AURA's advantage is most pronounced in dual-source localization: its $48.21\%$ DoA accuracy exceeds the strongest baseline, DSpAST-BAT$^{\dagger}$ ($37.90\%$), by $10.31$ points, and moving from single- to dual-source inputs costs AURA $25.4$ points against $35.6$ for BAT and $42.9$ for DSpAST-BAT$^{\dagger}$. On single-source inputs, AURA reaches $73.58\%$, above BAT ($72.28\%$) but below DSpAST-BAT$^{\dagger}$ ($80.84\%$). AURA also has the lowest single- and dual-source DER ($42.69\%$ and $64.72\%$), its relation direction BA ($68.86\%$) is above BAT's ($68.54\%$), and its relation distance BA ($70.65\%$) is $0.70$ points below BAT's ($71.35\%$).

These results require no dedicated spatial encoder training: AURA updates $64.63$M parameters, approximately $40\%$ of the more than $161$M in BAT and DSpAST-BAT$^{\dagger}$, and uses $1.94$M training samples across its parameter-update stages. Reorganizing and supervising a frozen general-audio codec representation thus supports competitive spatial understanding without a separately trained spatial feature space.

\begin{table}[!t]
    \centering
    \small
    \setlength{\tabcolsep}{4.5pt}
    \caption{
        Results on SpatialSoundQA at the end of each curriculum stage, in
        percent. Each stage is initialized from the previous one; ``Samples''
        is the cumulative number of training samples. A dash means the task
        has not been introduced at that stage and was not evaluated. The
        majority baseline answers the majority class on each axis and the
        best constant distance.
    }
    \label{tab:curriculum}
    \begin{tabular*}{\linewidth}{@{\extracolsep{\fill}}llcrrrrrr@{}}
        \toprule
        \textbf{Stage} & \textbf{Training data} & \textbf{Samples}
        & \multicolumn{2}{c}{\textbf{Single Source}}
        & \multicolumn{2}{c}{\textbf{Dual Source}}
        & \multicolumn{2}{c}{\textbf{Relation}} \\
        \cmidrule(lr){4-5}\cmidrule(lr){6-7}\cmidrule(lr){8-9}
        & & & \textbf{DoA} $\uparrow$ & \textbf{DER} $\downarrow$
        & \textbf{DoA} $\uparrow$ & \textbf{DER} $\downarrow$
        & \textbf{Dir. BA} $\uparrow$ & \textbf{Dist. BA} $\uparrow$ \\
        \midrule
        Random & \multicolumn{1}{c}{--} & --
        & 12.57 & 67.33 & 12.41 & 67.46 & 50.00 & 50.00 \\
        Majority baseline & \multicolumn{1}{c}{--} & --
        & 15.57 & 49.31 & 13.62 & 71.34 & 50.00 & 50.67 \\
        \midrule
        Stage 1 & single-source & 0.33M
        & 71.59 & 41.03 & \multicolumn{1}{c}{--} & \multicolumn{1}{c}{--} & \multicolumn{1}{c}{--} & \multicolumn{1}{c}{--} \\
        Stage 2 & + dual-source & 0.85M
        & 72.09 & 42.77 & 46.15 & 64.74 & \multicolumn{1}{c}{--} & \multicolumn{1}{c}{--} \\
        Stage 3 (AURA) & + relation questions & 1.94M
        & 73.58 & 42.69 & 48.21 & 64.72 & 68.86 & 70.65 \\
        \bottomrule
    \end{tabular*}
\end{table}

\begin{wraptable}[23]{r}{0.56\linewidth}
    \vspace{-0.6\baselineskip}
    \centering
    \small
    \setlength{\tabcolsep}{3pt}
    \caption{
        Effect of the temporal pooling scheme on the linear readability of
        the features. Reference rows are the majority-class baseline.
        Frame-level uses every frame as a probe row and averages the frame
        scores over the clip; global mean and global max pool each slot over
        the whole clip; AURA is the windowed, band-specific scheme ($w{=}4$,
        max for content-oriented and mean for acoustic-oriented slots)
        followed by averaging over windows, so that each clip is one vector.
        Protocol, data, and columns as in Table~\ref{tab:probe_readability}.
        The best value in each column is in bold and the second best is
        underlined.
    }
    \label{tab:pooling_probe}
    \begin{tabular*}{\linewidth}{@{\extracolsep{\fill}}lccccc@{}}
        \toprule
        & \multicolumn{4}{c}{\textbf{Direction} $\uparrow$}
        & \textbf{Dist.} \\
        \cmidrule(lr){2-5} \cmidrule(l){6-6}
        \textbf{Temporal summary}
        & DoA & L/R & F/B & U/D
        & MAE $\downarrow$ \\
        \midrule
        \rowcolor{black!8}
        \multicolumn{6}{@{}l}{\textit{Single source}} \\
        Reference
        & 15.6 & 50.9 & 50.9 & 59.0 & 1.249 \\
        Frame-level
        & 43.4 & \textbf{93.0} & 70.4 & 65.5 & 1.152 \\
        Global mean
        & \underline{51.2} & \textbf{93.0} & \underline{72.9} & 73.4 & 1.029 \\
        Global max
        & 50.3 & 92.2 & 72.2 & \underline{73.7} & \underline{1.014} \\
        \textbf{AURA}
        & \textbf{53.6} & \underline{92.9} & \textbf{74.4} & \textbf{75.0} & \textbf{1.010} \\
        \midrule
        \rowcolor{black!8}
        \multicolumn{6}{@{}l}{\textit{Dual source (queried source)}} \\
        Reference
        & 15.4 & 50.4 & 49.8 & 55.5 & 1.695 \\
        Frame-level
        & 25.5 & \textbf{72.7} & 60.8 & 58.4 & 1.645 \\
        Global mean
        & \underline{28.3} & 72.1 & \textbf{63.9} & 61.1 & \textbf{1.576} \\
        Global max
        & 28.1 & \underline{72.3} & \underline{63.7} & \underline{61.4} & \underline{1.579} \\
        \textbf{AURA}
        & \textbf{28.4} & 72.2 & \textbf{63.9} & \textbf{61.8} & \textbf{1.576} \\
        \bottomrule
    \end{tabular*}
\end{wraptable}

\paragraph{General-audio understanding.} We also evaluate general-audio understanding on three benchmarks excluded from training with the candidate-likelihood protocol above. Table~\ref{tab:general_audio} reports AURA accuracies of $44.50\%$, $34.80\%$, and $32.71\%$ on MMAU test-mini, MMAR base, and AIR-Bench Foundation, above BAT, DSpAST-BAT$^{\dagger}$, and OWL on all three, while the two general-purpose models, trained with large-scale audio instruction data, score higher and serve as references. The spatially adapted model thus generalizes better to general-audio understanding than the specialized spatial systems, alongside its performance on spatial tasks.

\FloatBarrier
\section{Analysis}
\label{sec:analysis}
\subsection{Curriculum Learning}
\label{sec:analysis:curriculum}
Table~\ref{tab:curriculum} reports performance on the tasks introduced up to each curriculum stage, with random and majority-class predictors as references. The first stage uses single-source data, the second adds dual-source data, and the third adds relation questions, followed by continuation on relation questions to obtain the final AURA model; each stage is initialized from the preceding model and retains the data of earlier stages.

Direction perception improves as more complex tasks are introduced: single-source DoA accuracy rises from $71.59\%$ after the first stage to $72.09\%$ after the second and $73.58\%$ for the final model, and dual-source DoA accuracy from $46.15\%$ to $48.21\%$ after relation questions and the subsequent continuation. Accumulated training thus preserves earlier direction perception while extending the model to more complex tasks.

\subsection{Temporal Pooling}
\label{sec:analysis:pooling}
AURA pools each slot over windows of $w{=}4$ frames, with max pooling for content-oriented and mean pooling for acoustic-oriented slots (Section~\ref{sec:aura_method:representation}). To examine what this band-specific choice preserves, we reuse the linear-probe protocol of Section~\ref{sec:rvq_analysis:probe} on the full six-slot representation and vary only the temporal summary: frame-level readout without pooling, global mean, global max, and AURA's windowed pooling followed by averaging over windows, so that every variant yields one vector per clip. Probes are fitted on the training split and scored on the full evaluation split as in Table~\ref{tab:probe_readability}; Table~\ref{tab:pooling_probe} reports the results.

Any clip-level summary reads out more than frame-level features, and the band-specific scheme gives the best overall readout. Frame-level readout is the weakest on every attribute except the already saturated left/right: in single-source scenes its DoA accuracy is $43.4\%$ against at least $50.3\%$ for the pooled variants, with a correspondingly larger distance error. Global mean and global max are nearly indifferent for direction (front/back $72.9\%$ versus $72.2\%$, up/down $73.4\%$ versus $73.7\%$), with slightly lower distance error for max ($1.014$~m versus $1.029$~m), while the windowed, band-specific scheme reaches the best DoA ($53.6\%$) and distance MAE ($1.010$~m) and is within $0.1$ points of the best left/right accuracy. Dual-source scenes show the same ordering with smaller margins, the three clip-level schemes lying within $0.3$ points of each other in DoA. The probe isolates what a single clip-level vector preserves; AURA's language model additionally receives the full sequence of windows.

\subsection{Slot Ablation}
\label{sec:analysis:slot_ablation}
\begin{wraptable}[25]{r}{0.56\linewidth}
    \vspace{-0.6\baselineskip}
    \vspace{-14.5pt}
    \centering
    \small
    \setlength{\tabcolsep}{3pt}
    \caption{
        Slot-wise ablation: zeroing each of the six slots in turn at
        inference in the final AURA model, on the single-source and
        dual-source SSQA evaluation sets (direction accuracies and DER in
        percent). The zeroed token stays in the sequence at its position. All
        six slots of AURA contribute noticeably to the final performance. The
        best value in each column is in bold and the second best is underlined.
    }
    \label{tab:slot_zero}
    \begin{tabular*}{\linewidth}{@{\extracolsep{\fill}}lccccc@{}}
        \toprule
        & \multicolumn{4}{c}{\textbf{Direction} $\uparrow$}
        & \textbf{Dist.} \\
        \cmidrule(lr){2-5} \cmidrule(l){6-6}
        \textbf{Zeroed slot}
        & DoA & L/R & F/B & U/D
        & DER $\downarrow$ \\
        \midrule
        \rowcolor{black!8}
        \multicolumn{6}{@{}l}{\textit{Single source}} \\
        \textbf{Full model}
        & \textbf{73.58} & \underline{93.95} & \textbf{86.84} & \textbf{85.29} & \textbf{42.7} \\
        w/o $L_S$
        & 60.27 & 85.43 & 81.04 & 78.44 & 53.7 \\
        w/o $L_A$
        & \underline{71.72} & 93.34 & 85.85 & \underline{84.03} & 43.4 \\
        w/o $R_S$
        & 64.29 & 88.56 & 83.60 & 79.64 & 51.0 \\
        w/o $R_A$
        & 70.87 & 92.96 & 85.66 & 83.42 & 44.0 \\
        w/o $S_S$
        & 51.94 & 69.58 & 78.26 & 77.62 & 45.6 \\
        w/o $S_A$
        & 71.49 & \textbf{94.05} & \underline{86.07} & 83.85 & \underline{42.9} \\
        \midrule
        \rowcolor{black!8}
        \multicolumn{6}{@{}l}{\textit{Dual source (queried source)}} \\
        \textbf{Full model}
        & \textbf{48.21} & \textbf{85.13} & \textbf{71.72} & \textbf{68.64} & \textbf{64.7} \\
        w/o $L_S$
        & 41.80 & 76.96 & 69.80 & 66.29 & 69.8 \\
        w/o $L_A$
        & 45.97 & 84.00 & 70.82 & 67.51 & \textbf{64.7} \\
        w/o $R_S$
        & 41.72 & 78.08 & 69.92 & 65.29 & 68.4 \\
        w/o $R_A$
        & 44.75 & 83.75 & 70.62 & 67.11 & 65.5 \\
        w/o $S_S$
        & 31.53 & 62.09 & 65.51 & 62.60 & \textbf{64.7} \\
        w/o $S_A$
        & \underline{46.33} & \underline{84.88} & \underline{71.47} & \underline{67.75} & \underline{65.0} \\
        \bottomrule
    \end{tabular*}
\end{wraptable}

To see which of the six slots the trained model relies on, we zero one slot at a time at inference in the final AURA model without retraining: the slot's tokens remain in the sequence with positions and attention mask unchanged, and only their input embeddings are set to zero in every frame. The model is evaluated with the official protocol on the single-source ($29{,}656$ questions) and dual-source ($21{,}280$ questions) evaluation sets, and each condition is compared with the unmodified model on matched questions with paired exact McNemar tests (Table~\ref{tab:slot_zero}).

\begin{figure}[!t]
    \centering
    \includegraphics[width=0.92\linewidth]{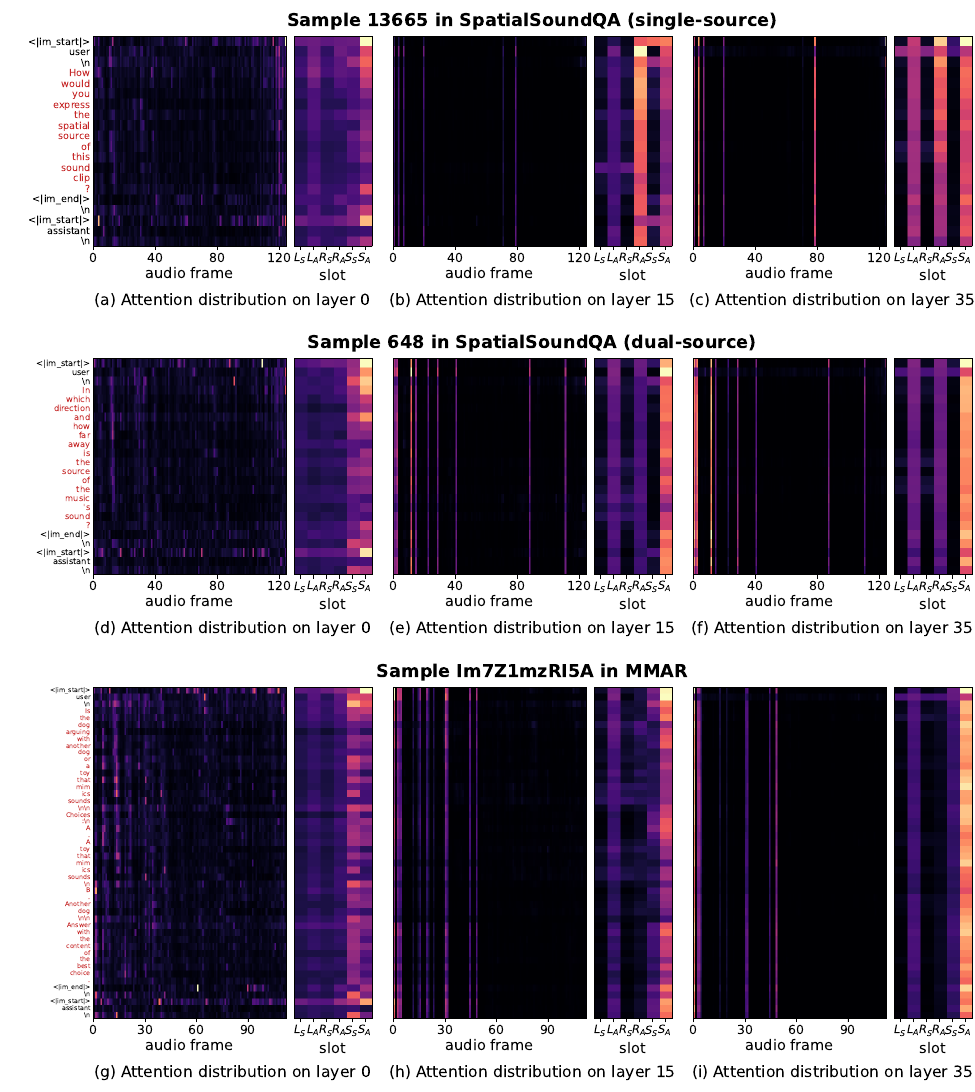}
    \caption{
        \textbf{Text-to-audio attention across layers.} Attention concentrates
        on more specific frames with depth and, depending on the question, is
        finally directed to different slots. Columns show the first (layer~$0$),
        an intermediate (layer~$15$), and the final (layer~$35$) transformer
        layer after prefill. Attention is averaged over heads. In each panel,
        the wide map sums attention over the six slots at each audio frame and
        the narrow map sums it over frames for each slot; rows are text tokens.
        Each map is normalized to its own maximum.
    }
    \label{fig:attn_grid}
\end{figure}
Every slot contributes to the final answer, and the content-oriented difference slot carries the largest share of lateral information. Zeroing any one of the six slots significantly lowers DoA accuracy on both evaluation sets (all $p<10^{-3}$). The largest effect comes from $S_S$: left/right accuracy falls from $93.95\%$ to $69.58\%$ on single-source and from $85.13\%$ to $62.09\%$ on dual-source questions, with DoA dropping by $21.6$ and $16.7$ points, and on dual-source questions both same-side and opposite-side left/right accuracy decrease ($-26.1$ and $-19.9$ points), so the slot serves both reading and binding. The content-oriented slots of the two ears have broader effects, as zeroing $L_S$ or $R_S$ degrades every direction axis and raises the single-source DER by $11.0$ and $8.3$ points. The acoustic-oriented slots complement them: zeroing $L_A$, $R_A$, or $S_A$ lowers DoA by $1.9$ to $3.5$ points, with the loss concentrated on front/back and up/down, the axes that depend on spectral cues, and on the opposite-side dual-source questions that require binding, so the two bands carry complementary parts of the spatial answer. Because $S_S$ is a deterministic function of $L_S$ and $R_S$, the information it carries remains present in the ear slots; that the trained model nevertheless relies on $S_S$ shows that the explicit difference slot is what makes lateral information accessible, consistent with the controlled experiment of Section~\ref{sec:information_asymmetry}, where adding the difference slots during training restored left/right reading.

\subsection{Attention Visualization}
\label{sec:analysis:attention}
We inspect text-to-audio attention during prefill, after the model has processed the audio and question but before it generates the first output token, averaged over heads and compared across the first (layer $0$), an intermediate (layer $15$), and the final (layer $35$) transformer layer. Figure~\ref{fig:attn_grid} shows examples answered correctly in the main evaluation, covering single-source, dual-source, and general-audio questions, with layers $0$, $15$, and $35$ from left to right; within a panel, the wide map aggregates attention over the six slots within each audio frame, and the narrow map aggregates over time for each slot. The first audio position, which acts as an attention sink \citep{attentionsink}, is excluded from the displayed maps, and the statistics in the next paragraph normalize each text row over the remaining audio positions.

Two patterns recur. First, attention concentrates on fewer audio frames at greater depth: across the inspected SSQA examples, the effective number of attended frames, the exponential of the frame-distribution entropy, decreases from approximately $115$ in layer $0$ to $16$ in layer $35$ out of $125$ frames per input, and general-audio examples show a similar trend. Second, the final-layer allocation over the six slots depends on the question. In the single-source example (c), attention is spread over the acoustic-oriented slots of both ears and of the difference stream ($L_A$, $R_A$, and $S_A$). In the dual-source example (f), where the queried source must be separated from the other one, the acoustic difference slot $S_A$ takes about half of the audio attention. In the MMAR example (i), which asks what is sounding rather than where, $S_A$ still dominates, but the content-oriented difference slot $S_S$ receives a visible share that is absent in the two spatial examples, while the right-ear acoustic slot $R_A$ is nearly ignored. The model thus routes attention to different slots for different questions rather than reading the six slots uniformly.

\section{Conclusion}
\label{sec:conclusion}
This paper has examined whether a general-audio codec trained without spatial supervision can support both identifying and locating sounds in multi-source scenes, a setting in which existing systems rely on a separately trained spatial feature space. A controlled layer-wise analysis of a frozen X-Codec showed that the spatial information is there. Shallow RVQ layers are shaped mainly by source content, deeper layers respond increasingly to interaural differences, room acoustics, and reverberation, and lateral information is recovered reliably only through the difference between the two channels; the same layer-wise trend holds on real binaural recordings. The presence of this information did not make it usable, however. A conventional interface trained with answer-only supervision remained at chance on left/right and read scene-level attributes without binding them to the queried source. We traced this failure to an audio--text information asymmetry, in which each clip contains two sources while the answer describes only one, together with an interface that leaves cross-channel comparison to the language model. In a controlled $2\times2$ experiment, scene supervision and an explicit interaural difference each restored left/right reading, and binding to the queried source improved substantially only when the two were combined. AURA builds on these observations. It aggregates the RVQ layers into content- and acoustic-oriented bands, adds the band-wise interaural difference to form a six-slot representation, and trains with scene and geometric supervision while the codec and the language model stay frozen. With about $40\%$ of the trainable parameters of dedicated systems, it exceeds the strongest of them by $10.31$ points in dual-source direction accuracy on SpatialSoundQA, achieves the lowest dual-source distance error rate, and scores higher than them on MMAU, MMAR, and AIR-Bench. Spatial and general-audio understanding therefore need not live in separate feature spaces. Once the spatial information in a frozen general-audio codec is made accessible and aligned with the text, that codec is sufficient for competitive multi-source spatial question answering.

\bibliography{references}
\bibliographystyle{plainnat}

\appendix
\section{Training Configuration}
\label{app:training}
Table~\ref{tab:app_training} lists the hyperparameters of the curriculum. Stage~1 first trains the projector alone with the language model and LoRA frozen, and every later row starts from the previous one and runs one epoch; training used four GPUs. The objective is the autoregressive language-model cross-entropy, with no auxiliary loss. All stages use AdamW ($\beta_1=0.9$, $\beta_2=0.999$, $\epsilon=10^{-8}$) without weight decay, a single learning rate for all trainable parameters, gradient-norm clipping at $1.0$, bfloat16 precision, and seed $42$. LoRA has rank $16$, $\alpha=32$, and dropout $0.05$, applied to the query, key, value, and output projections of attention and to the gate, up, and down projections of the feed-forward layers.

\begin{table}[!t]
    \centering
    \small
    \caption{Training stages of AURA. Each stage covers one epoch.}
    \label{tab:app_training}
    \begin{tabular}{@{}>{\raggedright\arraybackslash}p{0.25\linewidth}>{\raggedright\arraybackslash}p{0.69\linewidth}@{}}
        \toprule
        \textbf{Stage} & \textbf{Parameters} \\
        \midrule
        Stage 1, alignment &
        Trained: projector \newline
        Samples: $55{,}008$ single-source \newline
        Steps: $1{,}719$ \newline
        Global batch size: $32$ \newline
        Schedule: cosine with $100$ warmup steps \newline
        Learning rate: $10^{-4}$ \\
        \midrule
        Stage 1, joint &
        Trained: projector and LoRA \newline
        Samples: $278{,}784$ single-source \newline
        Steps: $17{,}424$ \newline
        Global batch size: $16$ \newline
        Schedule: cosine with $100$ warmup steps \newline
        Learning rate: $10^{-4}$ \\
        \midrule
        Stage 2 &
        Trained: projector and LoRA \newline
        Samples: $514{,}784$ single- and dual-source \newline
        Steps: $32{,}174$ \newline
        Global batch size: $16$ \newline
        Schedule: cosine with $100$ warmup steps \newline
        Learning rate: $10^{-4}$ \\
        \midrule
        Stage 3 &
        Trained: projector and LoRA \newline
        Samples: $872{,}312$ single-source, dual-source, and relation \newline
        Steps: $54{,}520$ \newline
        Global batch size: $16$ \newline
        Schedule: cosine with $100$ warmup steps \newline
        Learning rate: $10^{-4}$ \\
        \bottomrule
    \end{tabular}
\end{table}

\section{Dataset Composition}
\label{app:dataset}
Table~\ref{tab:app_dataset} gives the number of questions per type. Stage~1 uses the single-source rows, Stage~2 adds the dual-source rows, and Stage~3 adds the relation rows. Training and evaluation audio do not overlap. The non-binary relation questions are part of training but are not scored in the main results, whose binary accuracy uses the two binary relation types.

\begin{table}[!t]
    \centering
    \small
    \caption{Number of training and evaluation questions per type.}
    \label{tab:app_dataset}
    \begin{tabular*}{\linewidth}{@{\extracolsep{\fill}}llrr@{}}
        \toprule
        \textbf{Scene} & \textbf{Question type} & \textbf{Train} & \textbf{Eval} \\
        \midrule
        Single source & Direction and distance & 139,392 & 14,828 \\
        & Classification & 139,392 & 14,828 \\
        \midrule
        Dual source & Direction and distance & 118,000 & 10,640 \\
        & Classification & 118,000 & 10,640 \\
        \midrule
        Relation & Binary direction & 120,002 & 12,002 \\
        & Binary distance & 125,800 & 10,600 \\
        & Non-binary direction & 37,242 & 3,232 \\
        & Non-binary distance & 37,242 & 3,232 \\
        & Non-binary source & 37,242 & 3,232 \\
        \midrule
        Total & & 872,312 & 83,234 \\
        \bottomrule
    \end{tabular*}
\end{table}

\section{Question and Prompt Formats}
\label{app:prompts}
AURA receives the SSQA question unchanged in the Qwen3 chat template, with the audio tokens before the text; its target formats are shown in Fig.~\ref{fig:target_examples}, and Table~\ref{tab:app_questions} gives one original SSQA question per type with its answer. BAT, DSpAST-BAT$^{\dagger}$, and OWL use the prompt construction of their released code.

\begin{table}[!t]
    \centering
    \small
    \caption{Example questions and answers from the original SSQA dataset.}
    \label{tab:app_questions}
    \begin{tabular}{@{}>{\raggedright\arraybackslash}p{0.2\linewidth}>{\raggedright\arraybackslash}p{0.48\linewidth}>{\raggedright\arraybackslash}p{0.22\linewidth}@{}}
        \toprule
        \textbf{Type} & \textbf{Question} & \textbf{Answer} \\
        \midrule
        Direction and distance & How would you describe the location of the speech's sound in terms of direction and distance? & right, front, below; 0.5m \\
        Classification & Enumerate the sound occurrences in the audio clip. & accelerating, revving, vroom; car; vehicle \\
        Binary direction & Are you able to detect the orchestra's sound coming from the right and the cricket's sounds from the left? & Yes \\
        Binary distance & Is the sound of mechanisms further from you than male singing when considering the direct paths? & Yes \\
        Non-binary direction & Do you know if the tuning fork's sound comes from the left or right side of the sound made by the hum? & right \\
        Non-binary distance & What is the distance between the sound of the crumpling, crinkling and the sound of the music? & 2m \\
        Non-binary source & What can be heard on the behind side of the siren's audible sound? & music \\
        \bottomrule
    \end{tabular}
\end{table}

Qwen2-Audio and Qwen2.5-Omni receive the question followed by a format instruction, because they are not trained on SSQA. For direction and distance: ``Return exactly four fields and nothing else: left or right, front or behind, above or below; numeric distance in meters. Example: left, front, above; 2.5m.'' For classification and non-binary source questions: ``Answer only with the AudioSet sound-event label or labels. Separate multiple labels with semicolons. Do not explain.'' For binary relations: ``Answer exactly Yes or No and nothing else.'' For non-binary direction: ``Answer with exactly one direction word: left, right, front, behind, above, or below.'' For non-binary distance: ``Answer only with one numeric distance in meters, for example 2.5m.''

\section{Baseline Evaluation Details}
\label{app:baselines}
All models answer with single-path decoding: AURA with greedy search and at most $192$ new tokens ($384$ on relation questions), BAT, DSpAST-BAT$^{\dagger}$, and OWL with a beam size of $1$, and the two Qwen models with greedy search and at most $48$ new tokens.

\paragraph{BAT.} We use the released Spatial-AST encoder, eight-layer Q-Former, and LLaMA-2-7B LoRA checkpoint with its released inference code.

\paragraph{DSpAST-BAT$^{\dagger}$.} The released DSpAST Stage-3 encoder and LLaMA-2-7B stay frozen. We train an eight-layer Q-Former and LoRA (rank $8$, $\alpha=32$, dropout $0.05$) on the $872{,}312$ SSQA training questions for one epoch, which is $27{,}260$ steps at a global batch size of $32$, with a learning rate of $10^{-4}$ and $5{,}452$ warmup steps.

\paragraph{OWL.} We use the released Stage-1 checkpoint for single-source questions and the released Stage-2 checkpoint for dual-source questions, with the output mapping given in the Baselines paragraph of the main experiments.

\paragraph{Qwen2-Audio and Qwen2.5-Omni.} Both are evaluated zero-shot. The binaural input is reduced to one channel by a fractional delay-and-sum beam steered to the left at $30^\circ$.

\section{Layer-wise Similarity Across Codecs}
\label{app:codecs}
We repeat the interaural and placement comparisons of Section~\ref{sec:rvq_analysis:similarity} on two codecs trained for reconstruction without semantic supervision: DAC \citep{dac} at $16$~kHz with twelve layers, and EnCodec \citep{encodec} at $24$~kHz with thirty-two layers, whose input is upsampled from $16$~kHz. Inputs and pairs follow the same protocol. DAC quantizes in an $8$-dimensional factorized space and projects each codeword back to $1024$ dimensions before the residual subtraction; we use the projected vector as $\mathbf{c}_{t,i}$, and the $8$-dimensional entries give nearly the same curves (largest difference $0.044$, at L0). EnCodec codewords have $128$ dimensions and no output projection.

\begin{figure}[!t]
    \centering
    \includegraphics[width=\linewidth]{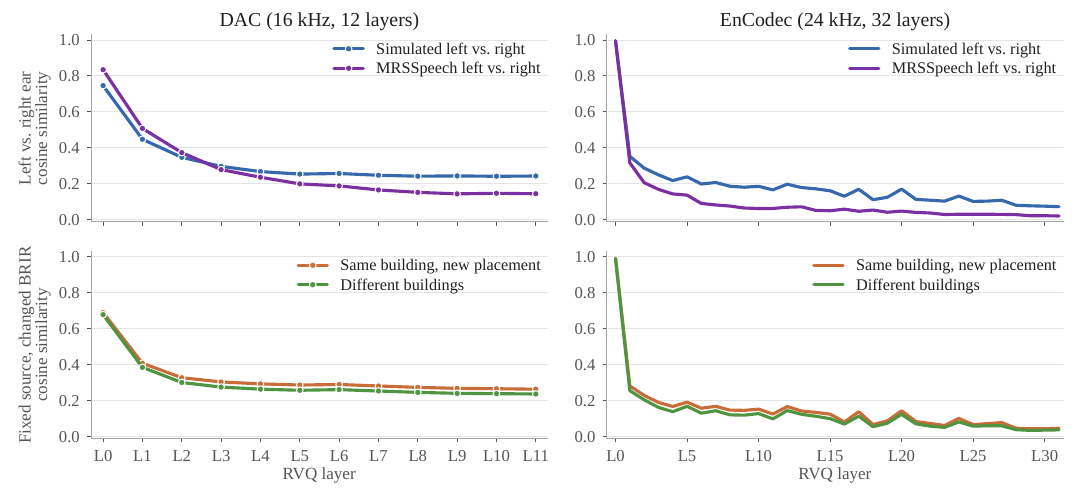}
    \caption{
        \textbf{Layer-wise similarity on reconstruction-only codecs.}
        Left: DAC; right: EnCodec.
        Top: interaural similarity for simulated (fixed source and BRIR) and real
        (MRSSpeech) binaural audio; the RVQ vectors diverge markedly as the layer
        index increases.
        Bottom: room variation: right-ear similarity when the same audio is
        rendered at another placement in the same building or with an impulse
        response from a different building; the similarity of the RVQ vectors
        still decreases clearly with depth.
        These comparisons reveal that the increasing sensitivity to interaural
        and room changes in deeper RVQ layers is not specific to X-Codec.
    }
    \label{fig:codec_similarity}
\end{figure}

Figure~\ref{fig:codec_similarity} shows that both trends appear on both codecs. On DAC, left/right similarity falls from $0.746$ at L0 to $0.243$ at L11 on simulated audio and from $0.835$ to $0.144$ on MRSSpeech; a new placement falls from $0.689$ to $0.262$ and a different building, which stays lower at every layer, from $0.677$ to $0.235$. DAC flattens after about L5. On EnCodec, from L0 to L31, the same comparisons fall from $0.995$ to $0.072$, $0.997$ to $0.020$, $0.991$ to $0.045$, and $0.991$ to $0.036$, with small local rises between layers. These two trends therefore do not require a semantic branch.

\section{Direction Probes on Real Binaural Recordings}
\label{app:mrsspeech_probe}

\begin{table}[!t]
    \centering
    \small
    \caption{
        Linear probes of spatial attributes on real binaural recordings
        (MRSSpeech), with the features of Table~\ref{tab:probe_readability}. Test
        speakers and every drama they appear in are excluded from training, and
        monaural features use the left ear. L/R+F/B requires both axes to be correct.
        Up/down has no entries because every annotated source lies on one side of the
        listener, so the axis has a single class. Reference rows are constant predictions
        without audio.
    }
    \label{tab:probe_mrsspeech}
    \begin{tabular*}{\linewidth}{@{\extracolsep{\fill}}lccccc@{}}
        \toprule
        & \multicolumn{4}{c}{\textbf{Direction} $\uparrow$}
        & \textbf{Dist.} \\
        \cmidrule(lr){2-5} \cmidrule(l){6-6}
        \textbf{Features}
        & L/R+F/B & L/R & F/B & U/D
        & MAE $\downarrow$ \\
        \midrule
        Reference & 50.6 & 61.6 & 56.1 & -- & 0.515 \\
        Monaural & 62.9 & 77.8 & 72.3 & -- & 0.496 \\
        Binaural & 67.2 & 84.5 & 74.6 & -- & 0.487 \\
        \bottomrule
    \end{tabular*}
\end{table}

We apply the recording-level probes of Section~\ref{sec:rvq_analysis:probe} to MRSSpeech \citep{mrsspeech}, using the same frozen X-Codec bands and classifier, with labels taken from the annotated source trajectories. The $5{,}501$ segments come from $30$ dramas recorded in one scene with one listener position; segments shorter than $1$~s are skipped and longer ones truncated to $20$~s. For each direction axis we remove segments whose source stays within $0.2$~m of the axis plane or crosses it, leaving $5{,}073$ left/right, $4{,}807$ front/back, and $4{,}494$ segments valid for both; distance is the mean radial distance of the trajectory, on all segments. Because a speaker tends to keep one position within a drama, each test speaker and every drama they appear in are excluded from training. We omit up/down and three-axis DoA because this dataset has no up/down contrast: the annotated source height ranges from $0.01$ to $0.67$~m and stays below the annotated ear height of $0.95$~m in every segment, so all sources fall on one side of the listener and the axis has a single class.

Table~\ref{tab:probe_mrsspeech} reports the results. Binaural features reach $84.5\%$ on left/right and $74.6\%$ on front/back against references of $61.6\%$ and $56.1\%$, and $67.2\%$ on both axes jointly against $50.6\%$. Distance error is $0.487$~m against a reference of $0.515$~m. As on SSQA, the gain from monaural to binaural features is largest on left/right.

\end{document}